\documentclass[%
 aip,
 amsmath,amssymb,
 reprint,%
]{revtex4-1}

\usepackage{graphicx}
\usepackage{dcolumn}
\usepackage{bm}

\usepackage[utf8]{inputenc}
\usepackage[T1]{fontenc}
\usepackage{mathptmx}
\usepackage{etoolbox}
\usepackage{xcolor}
\usepackage{placeins}
\usepackage{float}

\pdfpageattr{/Group << /S /Transparency /I true /CS /DeviceRGB>>}

\makeatletter
\def\@email#1#2{%
 \endgroup
 \patchcmd{\titleblock@produce}
  {\frontmatter@RRAPformat}
  {\frontmatter@RRAPformat{\produce@RRAP{*#1\href{mailto:#2}{#2}}}\frontmatter@RRAPformat}
  {}{}
}%
\makeatother

\begin{document}

\preprint{AIP/123-QED}

\title[Modeling of Injected Current Stream-Induced 3D Perturbations in Local Helicity Injection Plasmas]{Modeling of Injected Current Stream-Induced 3D Perturbations in \\Local Helicity Injection Plasmas}
\author{C. E. Schaefer}
 \altaffiliation{Author to whom correspondence should be addressed: \\ carolynschaefer126@gmail.com}
 \affiliation{Department of Nuclear Engineering and Engineering Physics, University of Wisconsin-Madison, Madison, Wisconsin 53706, USA}

\author{A. C. Sontag}
\affiliation{Department of Nuclear Engineering and Engineering Physics, University of Wisconsin-Madison, Madison, Wisconsin 53706, USA}%

\author{N. M. Ferraro}
\affiliation{Princeton Plasma Physics Laboratory, Princeton, New Jersey 08540, USA}%

\author{J. D. Weberski}
\altaffiliation{Present address: Helion Energy, Inc., 1415 75th St SW, Everett, Washington 98203, USA}
\affiliation{Department of Nuclear Engineering and Engineering Physics, University of Wisconsin-Madison, Madison, Wisconsin 53706, USA}%

\author{S. J. Diem}%
\affiliation{Department of Nuclear Engineering and Engineering Physics, University of Wisconsin-Madison, Madison, Wisconsin 53706, USA} 

\date{\today}

\begin{abstract}

Solenoid-free tokamak startup techniques are essential for spherical tokamaks and offer a pathway to cost reduction and design simplification in fusion energy systems. Local helicity injection (LHI) is one such approach, employing compact edge current sources to drive open field line current that initiates and sustains tokamak plasmas. The recently commissioned Pegasus-III spherical tokamak provides a platform for advancing this and other solenoid-free startup methods. This study investigates the effect of LHI on magnetic topology in Pegasus-III plasmas. A helical filament model represents the injected current, and the linear plasma response to its 3D field is calculated with M3D-C1. Poincaré mapping reveals substantial flux surface degradation in all modeled cases. The onset of overlapping magnetic structures and large-scale surface deformation begins near $\Psi_{N} \approx 0.37$, indicating a broad region of perturbed topology extending toward the edge. In rotating plasmas, both single-fluid and two-fluid models exhibit partial screening of the $n=1$ perturbation, with two-fluid calculations showing stronger suppression near the edge. In contrast, the absence of rotation leads to strong resonant field amplification in the single-fluid case, while the two-fluid case with zero electron rotation mitigates this amplification and preserves edge screening. Magnetic probe measurements indicate that modeling the stream with spatial spreading--representing distributed current and/or oscillatory motion--better reproduces measured magnetic power profiles than a rigid filament model. The results underscore the role of rotation and two-fluid physics in screening stream perturbations and point to plasma flow measurements and refined stream models as key steps toward improving predictive fidelity. 

\end{abstract}

\maketitle

\section{\label{sec:Intro}Introduction\protect}

Alternative approaches to plasma initiation that do not rely on a traditional Ohmic solenoid are critical for advancing the spherical tokamak (ST) \cite{Menard2019,Menard2022}. These solenoid-free startup techniques address limitations posed by the ST's narrow central column. They also offer the potential to reduce the cost and complexity of magnetically confined fusion energy systems more generally. 

DC helicity injection is a solenoid-free plasma startup and current drive technique that utilizes magnetic helicity input to grow and sustain the toroidal plasma current $I_{p}$ against resistive losses \cite{Jensen1984,Taylor1989}. In a tokamak, magnetic helicity $K = \int \textbf{A}\cdot\textbf{B}\; d^{3}x$ is a measure of the linkage between toroidal and poloidal magnetic fluxes. It is directly proportional to $I_{p}$ and the toroidal magnetic field $B_{T}$ (i.e., $K \sim I_{p}B_{T}$) \cite{Boozer1986}. The DC helicity injection approach creates a tokamak plasma by injecting $K$ via current injection at the plasma edge. In coaxial helicity injection (CHI), these edge current sources are large, coaxial electrodes to which voltage is applied, thereby driving current that initially flows along the open field lines that connect them \cite{Raman2001,Raman2007,Kuroda2022}. In contrast, local helicity injection (LHI) employs small, spatially-localized, non-axisymmetric sources at the plasma edge to inject current \cite{Ono1987,Darrow1990,Bongard2019}. 

LHI was studied extensively on the Pegasus Toroidal Experiment using plasma arc guns as the helicity injectors \cite{Garstka2006,Bongard2019}. This system demonstrated non-solenoidal plasma startup with high current multiplication, achieving $I_{p} \leq 0.25$ MA with injected current $I_{inj} \leq 8$ kA. The Pegasus-III experiment, a recently deployed upgrade to Pegasus, serves as a dedicated platform for advancing LHI and exploring other innovative non-solenoidal startup techniques \cite{Sontag2022}. The modeling inputs described in Section \ref{subsec:Inputs} are based on measurements from initial Pegasus-III experimental campaigns aimed at further advancing LHI and evaluating its performance at higher $B_{T}$.

Section \ref{sec:3DObservations} highlights that the magnetic geometry of an LHI plasma is inherently 3D. Nonlinear resistive simulations with NIMROD have shown that co-helicity stream reconnection events, in which driven flux ropes merge and release rings that accumulate poloidal flux, drive relaxation to a tokamak-like state and facilitate the sustainment of LHI \cite{OBryan2012,OBryan2014,OBryanPhD2014}. Experimental analysis has also shown that $I_{p}$ scaling in LHI plasmas aligns with trends predicted by stochastic field transport models \cite{Bodner2021}. In addition, an iterative method has been developed to correct for the systematic errors introduced by the 3D injected current when axisymmetry is assumed in equilibrium reconstructions \cite{Weberski2024}. The work presented here builds on this foundation by starting from a measurement-based equilibrium reconstruction and directly including the plasma response to the injected stream perturbation. This enables a focused investigation of how the streams modify the magnetic topology in experimentally relevant equilibria, the extent and nature of the chaotic edge region, and the significance of including the MHD plasma response in modeling this structure.

3D field effects are a critical consideration in both stellarator and tokamak configurations. In tokamaks, externally applied perturbations, such as resonant magnetic perturbations (RMPs), are known to alter edge magnetic topology, drive the formation of island chains and stochastic edge regions, and enhance transport, enabling control of edge-localized modes (ELMs) in high-confinement (H-mode) plasmas \cite{Evans2006, Suttrop2018}. Similarly, the 3D perturbation introduced by LHI edge currents can significantly impact edge structure, transport properties, and confinement. Understanding the effects of these 3D injected current structures is essential not only for studying LHI, but also for evaluating the quality of the LHI-initiated plasma that may be handed off to subsequent current drive mechanisms.

Furthermore, while RMP coils are typically stationary external components, the use of edge currents has been explored in some tokamaks as a potentially more robust alternative in reactor-relevant environments, where neutron damage may reduce coil longevity \cite{Rack2014,Hao2022}. Previous Pegasus experiments also found that modest LHI drive in H-mode plasmas reduced ELM activity without degrading confinement, suggesting that low-power LHI could serve as a tool for modifying edge stability through magnetic perturbations \cite{Thome2017}. Although Pegasus-III focuses on LHI as a startup and current drive method--with higher $I_{inj}$ levels than those used in ELM suppression studies--there is overlap in the modeling approaches and the magnetic topology effects of the perturbations. 

These 3D effects are shaped not only by the field structure generated by the current streams themselves but also by how the plasma responds to them. Studies on other tokamaks have demonstrated the importance of including the MHD plasma response when evaluating the impact of applied 3D perturbations \cite{Canal2017,Frerichs2023}. For example, modeling efforts have shown that the response can either amplify or shield components of the perturbation in ways that significantly influence the resulting magnetic topology. Therefore, capturing the plasma response is essential for understanding the 3D magnetic structure produced by the injected current streams in an LHI plasma. This motivates the response modeling carried out in this work.

This paper presents an initial study aimed at evaluating the impact of injected current structures in LHI plasmas. Section \ref{sec:Streams} reviews the existing understanding of how these streams relax into a tokamak-like configuration and highlights observations—based on both measurements and simulation—of the system’s inherently 3D nature. Section \ref{sec:Modeling} describes the modeling approach for the injected current, presents the experimental measurements used to inform inputs to the plasma response calculations, and provides a brief overview of the M3D-C1 code used in the analysis. Section \ref{sec:Results} presents the modeling results, comparing cases with and without plasma response, and further distinguishes between single- and two-fluid treatments, both with and without rotation. Section \ref{sec:Refine} proposes a method for refining the stream model using magnetic measurements from an insertable Hall probe. Finally, Section \ref{sec:Conclusions} concludes with a summary of the findings presented.

\section{\label{sec:Streams}Injected Current Streams in LHI Plasmas}

\subsection{\label{sec:Relaxation}Relaxation and Current Drive}

\begin{figure}
\includegraphics[width=\columnwidth]{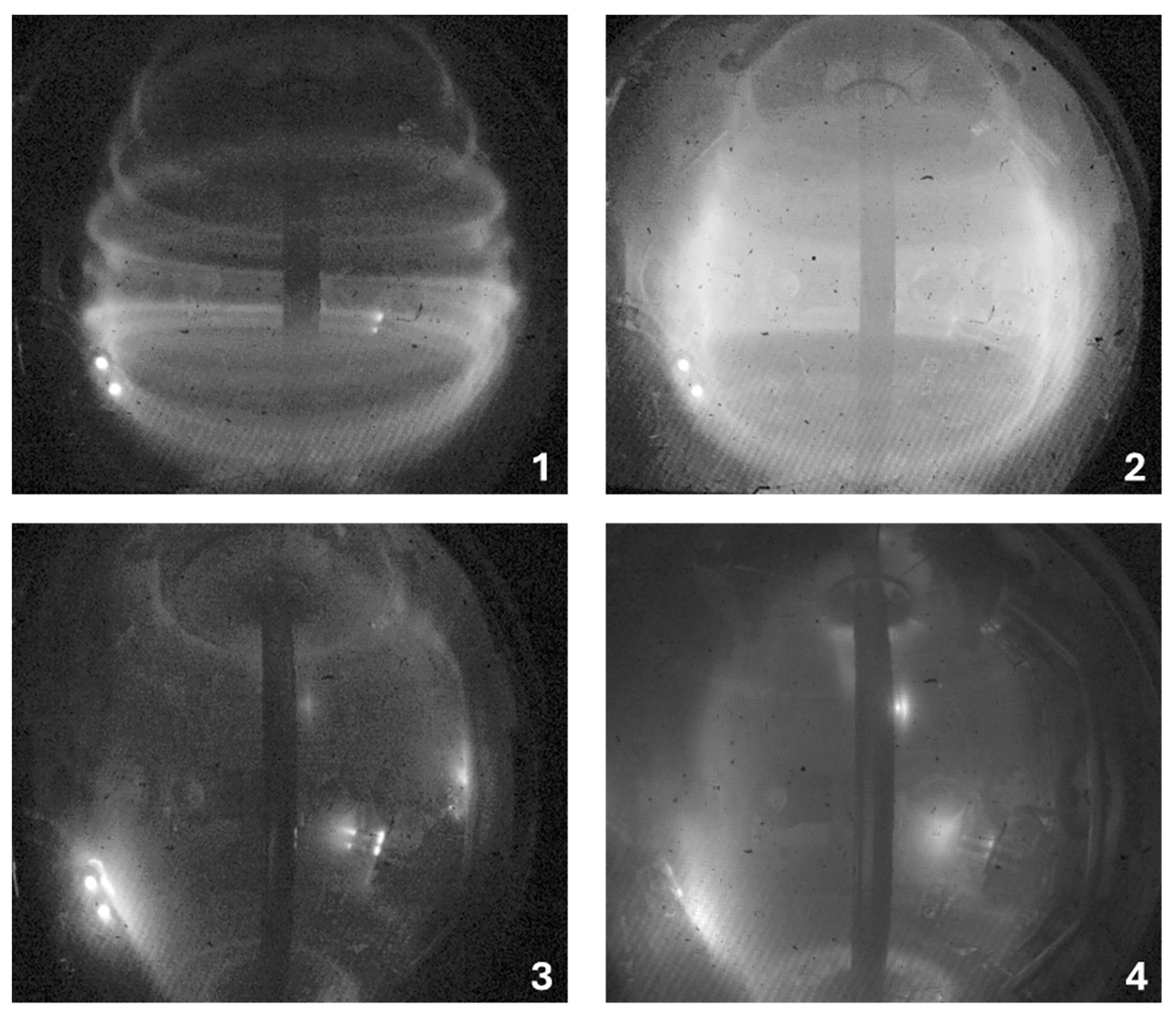}
\caption{\label{fig:LHI_Phases} Visible camera images of the startup and relaxation process during LHI on Pegasus-III: (a) injected current streams follow helical vacuum field lines; (b) streams become unstable and reconnect; (c) a tokamak-like plasma forms and is sustained by helicity input and inductive drive; (d) after injector shutoff, the perturbation dissipates and flux surfaces re-form, enabling handoff to other current drive systems.}
\end{figure}

The process by which the injected open field line current streams form a tokamak plasma during LHI occurs in multiple phases. The visible camera images shown in Figure \ref{fig:LHI_Phases} capture LHI startup and relaxation at various stages of the evolution. Using a washer-stack arc configuration, the process begins by applying a bias across an arc tube within the injector. Deuterium gas is then fed into the arc tube and ionized, thereby establishing an arc discharge current. The arc anode, which is the injection cathode, is biased negative relative to the vacuum vessel, which is grounded and acts as the injection anode. This potential difference causes an electron current $I_{inj}$ to be extracted from the arc plasma. 

As shown in the first frame of Figure \ref{fig:LHI_Phases}, the current streams initially follow the helical vacuum magnetic field geometry, making multiple passes around the torus. As $I_{inj}$ is increased, the streams become unstable, and adjacent stream passes attract one another. Under suitable conditions, these passes undergo magnetic reconnection, as illustrated in the second frame of Figure \ref{fig:LHI_Phases}. With sufficient $I_{inj}$, the streams eventually undergo a relaxation process, and a tokamak-like plasma is formed. This relaxed state is shown in the third frame of Figure \ref{fig:LHI_Phases}. Following this initial relaxation, the injectors continue inputting helicity to support the growth and maintenance of the tokamak-like plasma. In LHI with injectors mounted at the outboard edge, as on Pegasus-III, a combination of radial compression and dynamic shape evolution generates a significant non-solenoidal inductive voltage through poloidal field induction as the plasma expands radially inward \cite{Reusch2018,Barr2018}. Thus the discharge is sustained by both an inductive drive and the helicity drive directly from the LHI injectors. Once the injectors are shut off, the stream perturbation dissipates, and the magnetic flux surfaces rapidly heal, resulting in a tokamak plasma like the one shown in the fourth frame of Figure \ref{fig:LHI_Phases} that can be handed off to another current drive system, if available.

The LHI relaxation process redistributes the locally injected helicity across global scales. This relaxation is mediated by global physical mechanisms, including the requirement that magnetic helicity acts as a stronger invariant than magnetic energy. As a result, the magnetic topology evolves toward a minimum-energy "Taylor state," which facilitates the redistribution of currents and magnetic fields. In conjunction with these global mechanisms, magnetic fluctuation activity acts as the energy dissipation and current redistribution mechanism during the relaxation. Previous work presented a hypothesis for a relaxation and current drive mechanism motivated by observations of broadband ($f \sim 100$ kHz - 4 MHz) magnetic activity present during LHI \cite{Richner2022}. In this process, beam instabilities in the injected current streams drive Alfv$\acute{\text{e}}$nic turbulence, resulting in relaxation and current drive.

 These processes are thought to be active concurrently with the discrete, macroscopic stream reconnection events described above that also drive current. The role of macroscopic reconnection in relaxation and current drive in LHI is informed by previous work that simulated LHI on Pegasus using NIMROD \cite{OBryan2012,OBryan2014,OBryanPhD2014}. This work found that the initial tokamak-like plasma formation is due to an island-coalescence-like process by which consecutive helical passes of the injected current stream attract, undergo unstable kink-like oscillatory motion, and eventually reconnect. This produces a discrete axisymmetric current ring that separates from the injected current filament and freely diffuses inward. This ring formation process repeats at a rapid rate, and the accumulated poloidal flux within the confinement volume forms the initial tokamak-like state. This continues into the tokamak sustainment phase, and synthetic diagnostic measurements from the simulation show bursts of $n=1$ activity on the LFS magnetics that are linked to coherent motion and complete reconnection of the streams--specifically, their destabilization, merging and subsequent reestablishment. 
 
 In addition to the agreement between visible camera images and the NIMROD simulations during the initial formation phase, bursty $n=1$ activity is observed on LFS magnetic measurements during LHI operation even after the initial relaxation when the discrete streams are no longer discernible in visible imaging. This activity typically occurs at frequencies of 20-50 kHz and is well-characterized as a line-tied kink instability of the injected current streams \cite{HinsonPhD2015,BarrPhD2016,Reusch2018}. According to this interpretation, spatial displacements of the streams due to their line-tied kinking motion can bring neighboring passes close enough to attract one another and reconnect via the island-coalescence-like events described by NIMROD. Experimental observations also indicate that the injected current streams persist as coherent structures in the plasma edge region throughout the tokamak-like phase of LHI, with their kink motion being the source of the $n=1$ bursts. These bursts can coincide with increases in $I_{p}$, providing further experimental evidence that this reconnection process is a current drive mechanism that is active during LHI, at least intermittently \cite{Weberski2025}.
 
 Other observations suggest that this macroscopic reconnection is not the sole current drive mechanism present during LHI. Anomalously high ion heating due to reconnection activity was found to correlate more strongly with high-frequency (200-400 kHz) magnetic activity than with the lower-frequency $n=1$ bursts, indicating a possible role for small-scale reconnection \cite{Burke2017}. Additionally, sustained current drive has been observed in regimes where the $n=1$ activity is absent or strongly suppressed \cite{Reusch2018}. These observations challenge the idea that large-scale reconnection alone is responsible for current drive during LHI. Still, the persistence of coherent stream structures in the edge region--which underpins the macroscopic reconnection mechanism--supports the injected current stream model that is employed in this work and described in Section \ref{sec:CurrentStreamModel}. 

\subsection{\label{sec:3DObservations}Observations of Current Stream-Induced Non-Axisymmetric Magnetic Perturbation}

The LHI plasma that exists after relaxation is described as ``tokamak-like'' to reflect that the majority of the magnetic energy exists in the axisymmetric $n=0$ mode, as shown by prior NIMROD simulations of LHI \cite{OBryan2014}. This terminology reflects the assumption that the magnetic structure resembles that of a tokamak plasma when averaged over the toroidal direction. The tokamak-like state aligns with experimental results from LHI studies conducted not only on Pegasus \cite{Eideitis2007,Battaglia2009,Barr2018}, but also on CDX \cite{Ono1987} and VEST \cite{Park2022}. However, experimental observations indicate that the injected current streams remain as discrete structures (rather than diffusing into an axisymmetric distribution) in the plasma edge region throughout the tokamak-like phase \cite{HinsonPhD2015,BarrPhD2016,Reusch2018,Schaefer2024}. 

\begin{figure}
\includegraphics[width=\columnwidth]{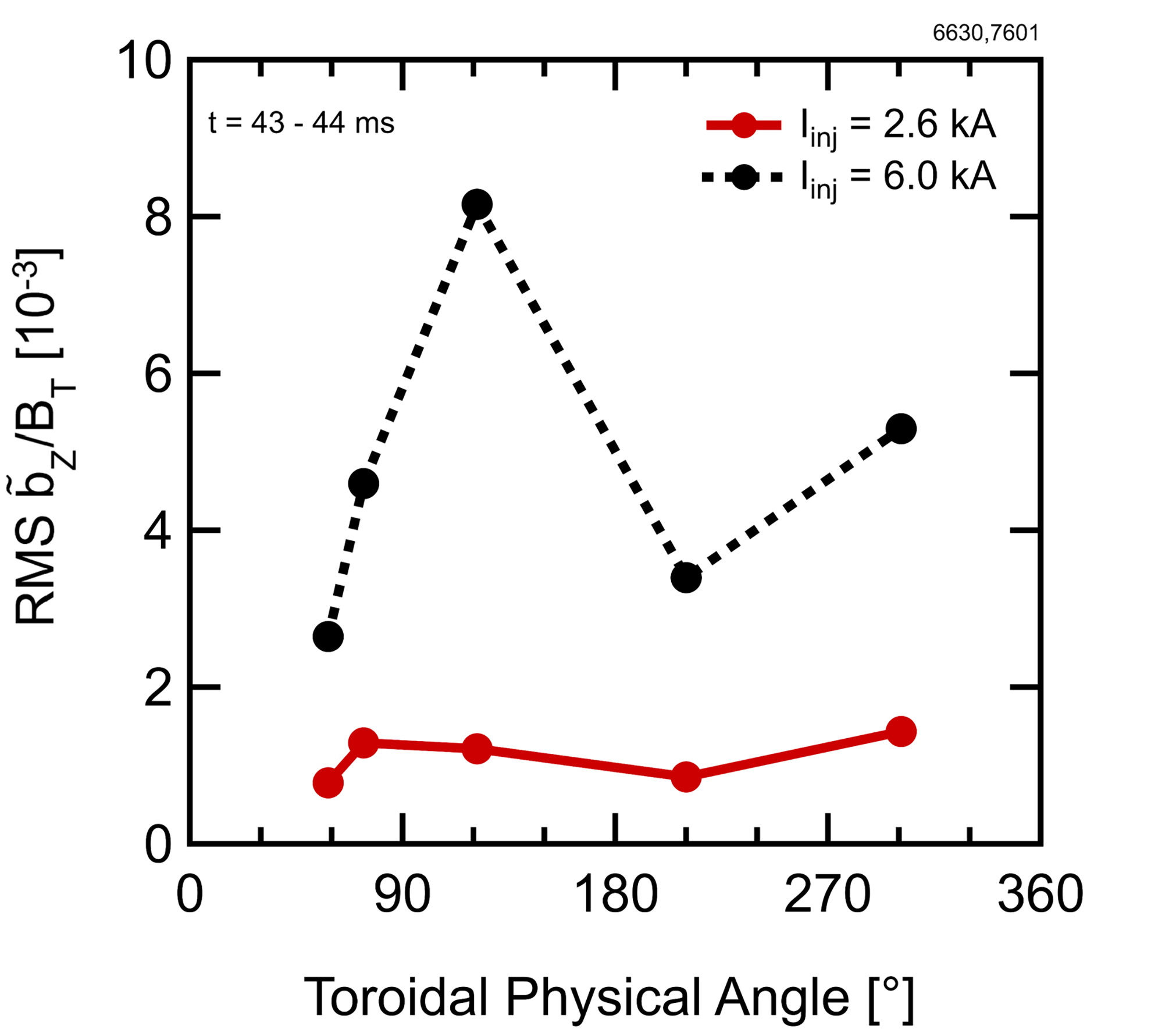}
\caption{\label{fig:Btilde} Toroidal dependence of magnetic activity, measured via the RMS of vertical magnetic field perturbation $\tilde{b}_{Z}$, normalized to the equilibrium toroidal magnetic field $B_{T}$. Data were collected from five magnetic pickup coils distributed toroidally near the outboard midplane. The red solid line corresponds to a discharge with $I{\text{inj}} = 2.6$ kA, and the black dashed line corresponds to a discharge with $I_{\text{inj}} = 6.0$ kA.}
\end{figure}

Furthermore, this stream structure is fundamentally 3D. Figure \ref{fig:Btilde} shows the toroidal dependence of magnetic activity from five poloidal magnetic field pickup coils distributed toroidally around the outboard vacuum vessel wall just below midplane, quantified using the root-mean-square (RMS) value of the vertical magnetic field perturbation $\tilde{b}_{Z}$ normalized to the equilibrium toroidal magnetic field $B_{T}$. The data shown is from two different discharges, each initiated with just one of the two injector arrays installed on Pegasus-III--one with a total $I_{inj}$ of 2.6 kA and the other with 6.0 kA. Notably, the discharge with higher $I_{inj}$ exhibits both larger magnetic fluctuations and a greater degree of relative toroidal variation, defined here as the difference between maximum and minimum values across toroidal positions normalized to the average. This suggests that both the magnitude and the non-axisymmetric character of the stream-driven perturbation increase with injected current level. In the lower-current case only one injector was operated, whereas the higher-current case used both injectors in the array, so stream-stream interactions may also contribute to the enhanced asymmetries observed.

As shown, the relative amplitude of the magnetic perturbation created by the injected current streams is on the order of $10^{-3}$ for these discharges, which is typical for LHI. Here, $\tilde{b}_{Z}$ denotes the time-varying component of the vertical magnetic field, obtained by subtracting the equilibrium field from the total signal. The quantity plotted in Figure \ref{fig:Btilde} represents the relative amplitude of magnetic fluctuations and serves as an indicator of the MHD activity associated with the dynamics of the injected current streams and other non-axisymmetric behavior in the plasma. This differs from the static perturbation field typically referenced in discussions of error fields or applied RMPs \cite{Park2023}. Still, it provides a useful order of magnitude estimate of the relative strength of the perturbed magnetic field, comparable in form to $|\delta\boldsymbol{B}|/|\boldsymbol{B}_{0}|$, which is known to deform magnetic surfaces and degrade performance at levels as small as $10^{-4}$ on other tokamaks. The LHI perturbation, therefore, is expected to deform magnetic surfaces and potentially contribute to the formation of a chaotic edge region. 

\begin{figure}
\includegraphics[width=\columnwidth]{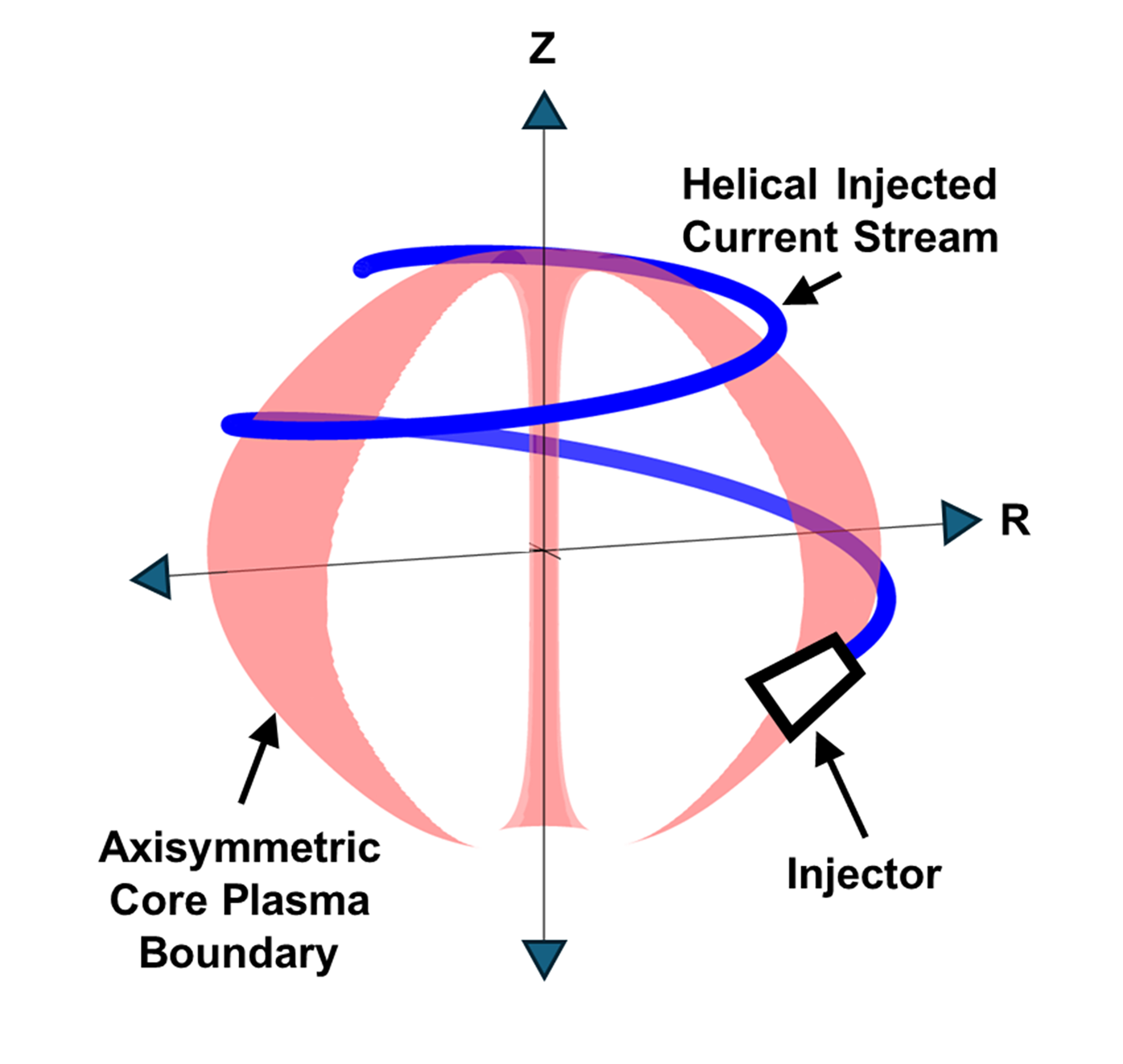}
\caption{\label{fig:Current_Stream_Model} Modeled system including an axisymmetric core plasma from equilibrium reconstruction (red surface), surrounded by a helical injected current stream (blue) that originates at an injector on the plasma boundary (white marker).}
\end{figure}

Previous work on Pegasus also demonstrated significant toroidal asymmetries in magnetic activity during LHI discharges \cite{BarrPhD2016,RichnerPhD2021}. One study systematically assessed the impact of the 3D perturbation caused by the injected current streams on equilibrium reconstruction of the resulting tokamak-like plasma \cite{Weberski2024}. In that analysis, the stream geometry was modeled by assuming an axisymmetric tokamak core surrounded by helical current streams--the same method described in Section \ref{sec:CurrentStreamModel}. A depiction of this system is shown in Figure \ref{fig:Current_Stream_Model}. The resulting non-axisymmetric magnetic fields were computed via a Biot-Savart approach and evaluated at the locations of standard reconstruction diagnostics, revealing perturbations that exceeded typical measurement uncertainties and highlighting the potential for systematic errors when axisymmetry is assumed. This further supports the inherent 3D nature of the streams, as the non-axisymmetric magnetic contributions are large enough to measurably impact diagnostic signals.

\section{Modeling Approach}
\label{sec:Modeling}

The modeling approach adopted in this study treats the injected current stream analogously to a static coil, producing a prescribed 3D magnetic perturbation that interacts with an otherwise axisymmetric Pegasus‐III equilibrium. The model comprises three main components: (1) a current stream model representing each injector as a helical filament whose geometry is determined from field‐line tracing of the reconstructed axisymmetric equilibrium; (2) an axisymmetric equilibrium reconstruction of the post‐relaxation LHI plasma using magnetic diagnostic constraints, supplemented by Thomson scattering measurements of electron temperature and electron density for defining the plasma state; and (3) a non-axisymmetric plasma response model using M3D‐C1 to calculate the linear, time‐independent single‐ or two‐fluid response to the imposed stream perturbation. 

\subsection {Current Stream Model}
\label{sec:CurrentStreamModel}

A model for the injected current similar to that shown in Figure \ref{fig:Current_Stream_Model} was employed. Each injected current stream is represented as a single filament originating at an injector aperture near the outboard plasma edge, following a helical magnetic field line upward toward the upper divertor. The filament is truncated at the upper boundary of the axisymmetric core plasma, where the field line would otherwise wrap downward along the LFS edge of the plasma. This truncation is ad hoc, as it does not explicitly specify a return location for the injected current, but it is consistent with experimental observations of stream localization to the outboard edge, stream coherence length estimates based on kinetic Alfvén wave damping, and the presence of a grounded conducting plate at the top of the vessel that likely serves as a current return path \cite{Weberski2024,Burke2017,Reusch2018,BarrPhD2016,PerryPhD2018,PachicanoMasters2018}. The helical stream geometry was determined by tracing field lines derived from the axisymmetric component of the magnetic field, as specified by the equilibrium reconstruction described in Section \ref{subsec:Inputs}. The Pegasus-III plasma discharge analyzed here used two injector arrays with two injectors each, resulting in four modeled helical filaments. For simplicity, each stream is modeled as a filament of negligible cross-section that carries a uniform current $I_{inj}$ along its length. The total $I_{inj}$ for this plasma was $\sim12$ kA, with each of the four streams carrying $\sim3$ kA. A Biot–Savart calculation is used to determine the 3D magnetic field produced by the helical current filament at all locations within the toroidally extended M3D-C1 grid. This field is then read into M3D-C1 and incorporated into the plasma response calculations, as described in Section \ref{subsec:M3DC1}.

Experimental observations have indicated that the stream structure is well approximated by a line-tied kink geometry anchored at the injector, and that the associated $n=1$ magnetic activity described in Section \ref{sec:Relaxation} is radially localized, peaking in the plasma edge region \cite{HinsonPhD2015,BarrPhD2016,Burke2017,RichnerPhD2021,Schaefer2024}. However, modeling the injected current streams as simple helical filaments that follow static magnetic field lines likely oversimplifies a more complex, dynamic current distribution. Nonetheless, given both the helical structure visible in fast camera images prior to relaxation and evidence of persistent 3D structure afterward, the simple helix remains a useful model for gaining an initial understanding of how the injected current impacts the equilibrium magnetic topology of a relaxed LHI plasma.

\subsection{Equilibrium Reconstruction and Thomson Scattering Measurements}
\label{subsec:Inputs}

\begin{figure}
\includegraphics[width=\columnwidth]{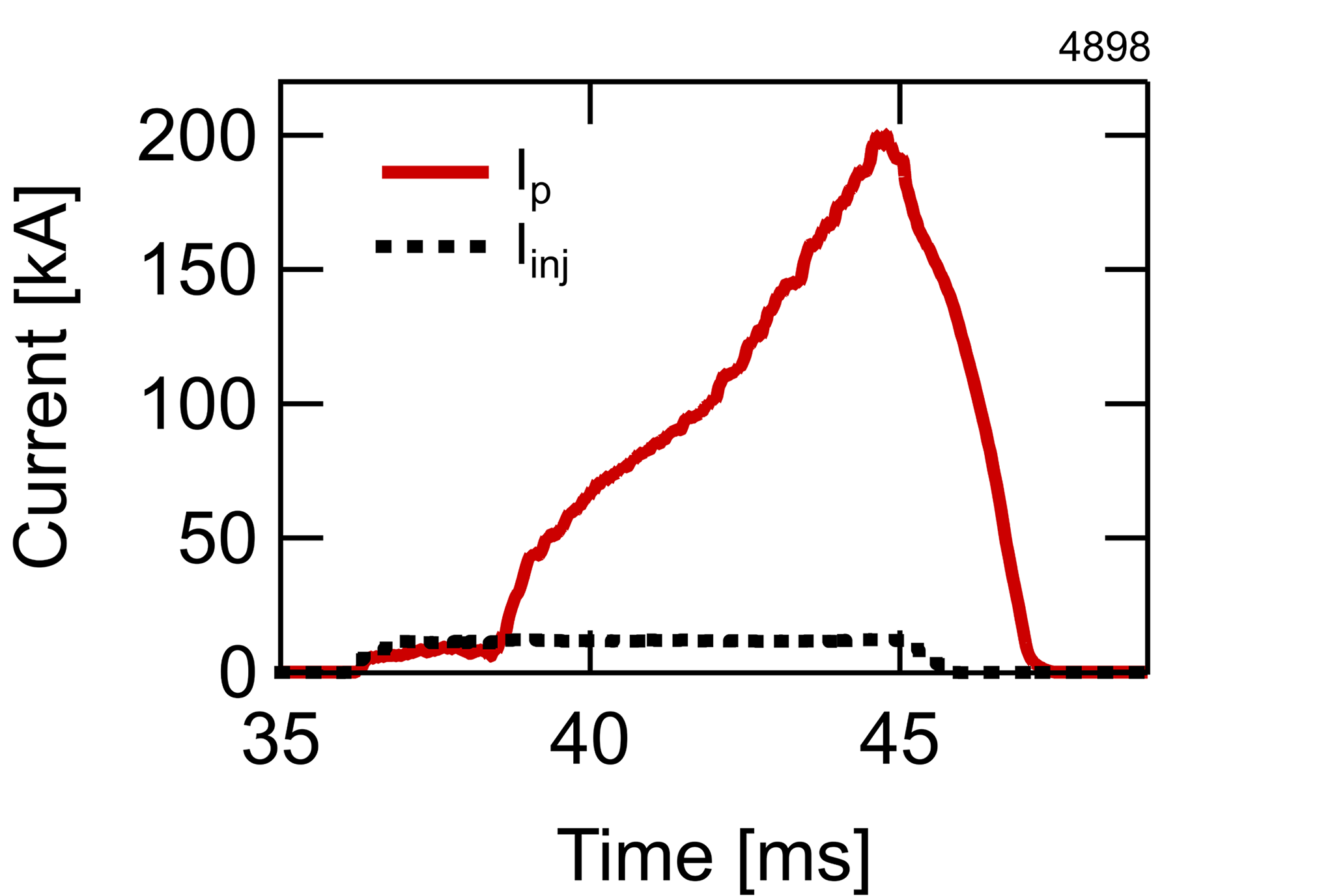}
\caption{\label{fig:Ip_and_Iinj} Time evolution of the plasma current ($I_{p}$, solid red line) and injected current ($I_{\text{inj}}$, dashed black line) for the Pegasus-III plasma discharge reconstructed for this study. The $I_{inj}$ shown represents the sum over all four injectors.}
\end{figure}

\begin{figure}
\includegraphics[width=\columnwidth]{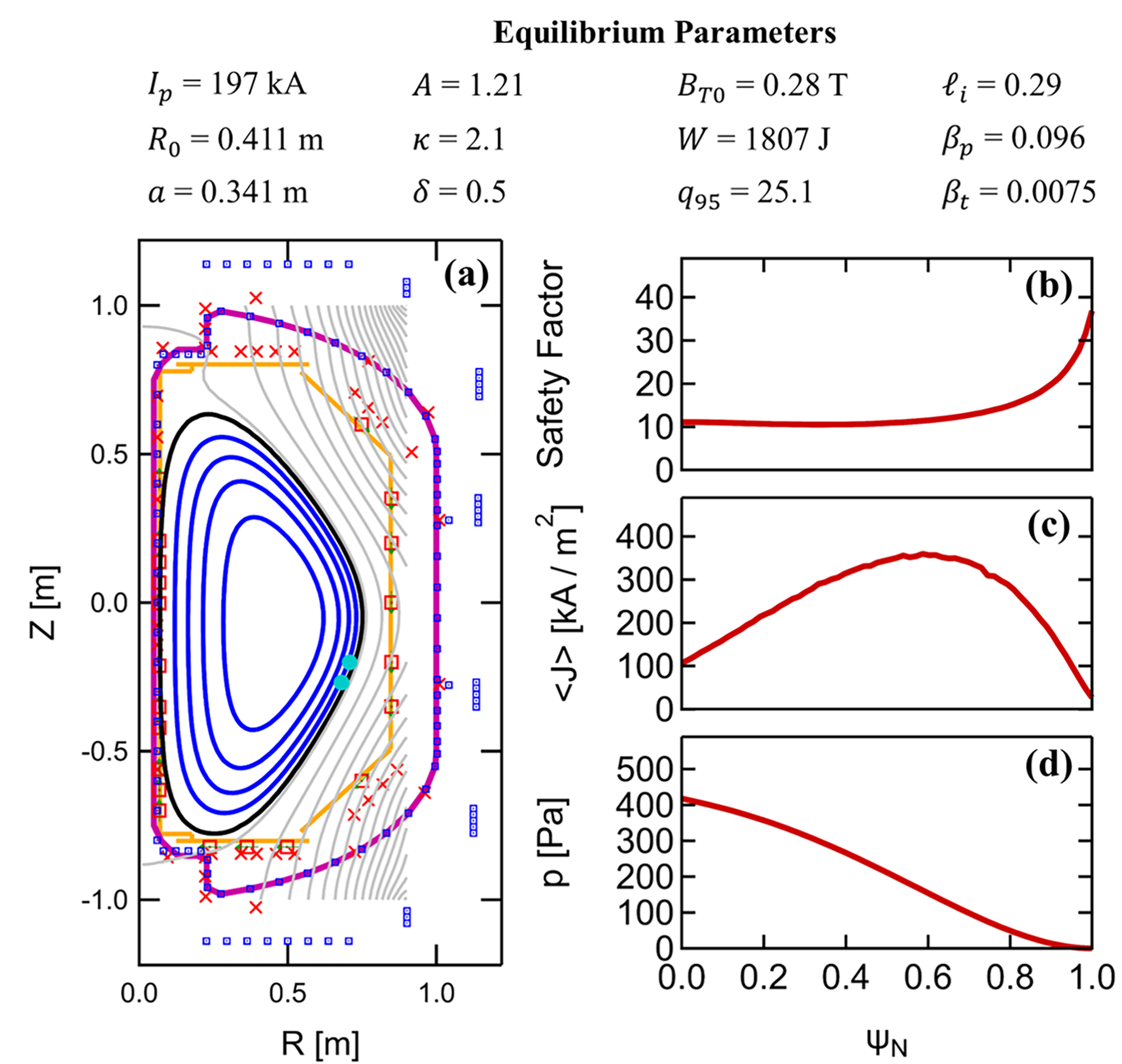}
\caption{\label{fig:Equilibrium_Reconstruction_with_Profiles} Overview of the reconstructed equilibrium for a Pegasus-III LHI plasma, including: (a) poloidal magnetic flux surfaces and locations of magnetic diagnostics used as constraints; and (b–d) profiles versus normalized poloidal flux $\Psi_{N}$ of the safety factor, volume-averaged toroidal current density $\langle J \rangle$, and plasma pressure $p$.}
\end{figure}

A Grad-Shafranov equilibrium was reconstructed to represent the axisymmetric component of the tokamak-like plasma that exists after relaxation of an LHI discharge. This reconstruction was obtained using the KFIT free-boundary equilibrium code \cite{Sontag2008,Bongard2011,Bongard2014} and constrained by measurements from the new magnetic diagnostic suite for Pegasus-III \cite{Reusch2024}. Included in these diagnostic constraints were 25 poloidal flux loops internal to the vessel wall, 16 poloidal flux loops external to the vessel wall, and 21 poloidal magnetic field pickup coils. The plasma reconstructed is a Pegasus-III plasma with $B_{T} \sim 0.3$ T, $I_{p} \sim 200$ kA driven by two arrays of two injectors (four total) located at the $(R,Z)$ positions indicated by the light blue dots in Figure \ref{fig:Equilibrium_Reconstruction_with_Profiles}a and toroidally separated by $180^{\circ}$. The $I_{p}$ and $I_{inj}$ evolutions for this plasma discharge are shown in Figure \ref{fig:Ip_and_Iinj}. 

An overview of the equilibrium solution is provided in Figure \ref{fig:Equilibrium_Reconstruction_with_Profiles}, showing a summary of the equilibrium parameters; reconstructed poloidal flux surfaces and diagnostic locations (\ref{fig:Equilibrium_Reconstruction_with_Profiles}a); and profiles versus normalized poloidal flux $\Psi_{N}$ of the safety factor $q$ (\ref{fig:Equilibrium_Reconstruction_with_Profiles}b), volume-averaged toroidal current density $\langle J \rangle$ (\ref{fig:Equilibrium_Reconstruction_with_Profiles}c), and plasma pressure $p$ (\ref{fig:Equilibrium_Reconstruction_with_Profiles}d).  The off-midplane magnetic axis reflects a programmed vertical field imbalance. The high, relatively flat central $q$ with a sharp edge rise is typical of low-aspect-ratio plasmas and consistent with Pegasus-III operation. The hollow $\langle J \rangle$ profile and peaked $p$ profile are also characteristic of LHI discharges \cite{Bodner2021}.

The equilibrium solution was exported to an EQDSK file, a standard format that provides the poloidal flux function $\Psi$ on an $(R,Z)$ grid along with 1D profiles of pressure (as shown in Figure~\ref{fig:Equilibrium_Reconstruction_with_Profiles}d), the poloidal current function, and $q$. M3D-C1 reads this EQDSK file and re-solves the equilibrium on its own finite element mesh. The resulting axisymmetric equilibrium forms the basis for subsequent linear MHD response calculations.

\begin{figure}
\includegraphics[width=\columnwidth]{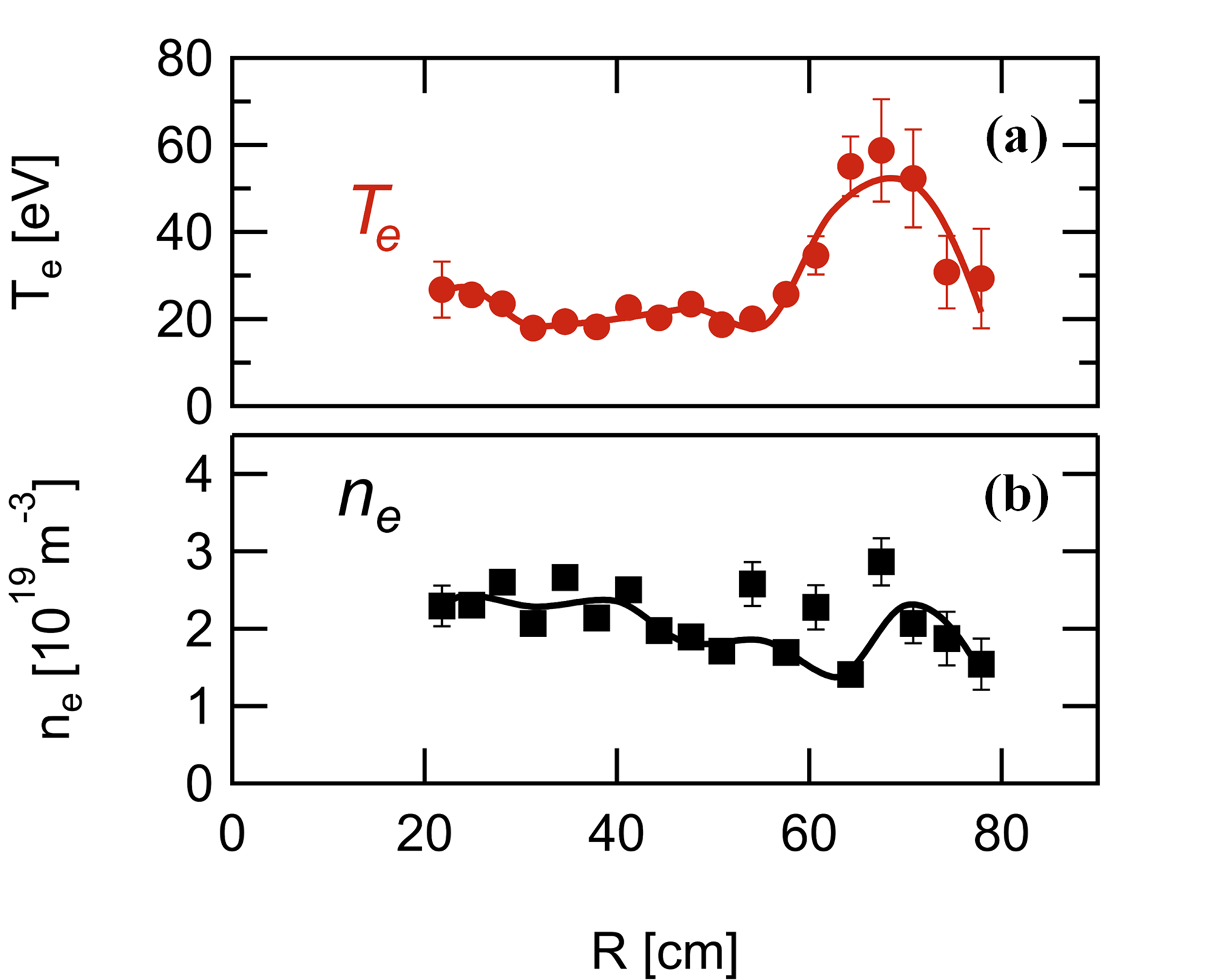}
\caption{\label{fig:ThomsonScatteringProfiles} Thomson scattering measurements of (a) electron temperature $T_{e}$ and (b) electron density $n_{e}$ for the Pegasus-III LHI plasma shown in Figure \ref{fig:Ip_and_Iinj} and reconstructed in Figure \ref{fig:Equilibrium_Reconstruction_with_Profiles}. Lines between points are smoothed splines for visual guidance.}
\end{figure}

Thomson scattering measurements of electron temperature $T_{e}$ and electron density $n_{e}$ corresponding to this same Pegasus-III plasma are shown in Figures \ref{fig:ThomsonScatteringProfiles}a and \ref{fig:ThomsonScatteringProfiles}b, respectively. The lines between points are smoothed splines for visual guidance \cite{Reinsch1967}. The hollow $T_{e}$ profile is consistent with typical Pegasus-III LHI observations \cite{Tierney2024}, although operation at this higher $B_{T}$ level compared to Pegasus may be expected to yield more peaked profiles \cite{Bodner2021}. The $n_{e}$ profile is similarly typical for Pegasus-III, though higher, more peaked values have been achieved with increased fueling \cite{Tierney2024}.  While these kinetic profiles are not constraints in the equilibrium reconstruction shown in Figure \ref{fig:Equilibrium_Reconstruction_with_Profiles}, they are supplied to M3D-C1 as fundamental inputs in both single- and two-fluid simulations and are used to derive auxiliary quantities.

\subsection{Plasma Response Physical Model}
\label{subsec:M3DC1}

The plasma response to the 3D field generated by the injected current streams in LHI plasmas is computed using the M3D-C1 code, a finite element code that solves the extended MHD equations\cite{Jardin2012,Ferraro2012,Ferraro2023}. M3D-C1 employs $C^{1}$-continuous, unstructured, triangular elements with reduced-quintic basis functions that form a mesh in the poloidal plane. For linear calculations, the toroidal direction is discretized spectrally using a single toroidal Fourier mode $n$. 

Here, the code is used to model the linear, time-independent response of the plasma to the stream perturbation. Both single-fluid and two-fluid models are considered. In single-fluid calculations, the terms in the extended MHD equations that are proportional to the normalized collisionless ion skin depth are omitted--specifically, the Hall term and the parallel electron pressure gradient term in the generalized Ohm's law. Previous measurements on Pegasus have indicated that ion temperature is greater than electron temperature in LHI plasmas due to magnetic reconnection activity \cite{Burke2017}, suggesting conditions where two‐fluid effects may be significant for accurately capturing the plasma response. Kinetic profiles of $T_{e}$ and $n_{e}$ are provided as inputs in both models to define the equilibrium plasma state and compute derived quantities. 

\section{Modeling Results}
\label{sec:Results}

The linear calculations presented in this section were performed for toroidal mode numbers $n=1$ through $n=8$, where $n$ corresponds to an independent toroidal harmonic of the perturbation induced by the injected current streams. Calculations at higher $n$ were limited by the spatial resolution of the computational mesh. The response calculations indicate non-negligible resonant fields across $n=1-8$, with amplitudes that exhibit distinct radial dependence and that span roughly an order of magnitude when considered across the different harmonics and cases. The full set of resulting field amplitudes is shown in Appendix \ref{sec:appendix}. Contributions from all of these harmonics are summed in the Poincaré plots presented in this section, providing a visualization of the total perturbed magnetic topology. While no single harmonic is assumed to dominate a priori, we focus detailed comparisons on the $n=1$ response because it is both experimentally prominent and physically relevant. As discussed in Section \ref{sec:Streams}, LHI plasmas consistently exhibit an $n=1$ spectral peak associated with macroscopic stream motion and reconnection. This is distinct from an intrinsic internal mode and therefore has an origin that differs from the linear resonances computed here, but its prominence makes $n=1$ a natural experimental reference case. In addition, low-$m$, $n=1$ activity has been observed in Ohmic plasmas on Pegasus, consistent with a genuine tearing-like instability in the absence of streams \cite{Garstka2003,Sontag2008,RichnerPhD2021,Richner2022}. The observation of such modes under Ohmic conditions demonstrates that the $n=1$ harmonic is physically significant beyond the LHI drive. Moreover, LHI regimes with equilibrium properties similar to those of Ohmic plasmas on Pegasus may likewise support this internal $n=1$ mode.

Comparisons are made between the ``vacuum'' field perturbation generated by the injected current stream, calculated using the Biot–Savart law, and the total perturbation that includes the plasma response. Additional comparisons are made between single-fluid and two-fluid calculations, with particular attention given to how variations in the toroidal rotation profile affect the plasma response. The influence of the $T_{e}$ profile is also briefly discussed.

\subsection{Single-fluid}

The single-fluid plasma response was calculated using smoothed spline fits derived from the measured $T_{e}$ and $n_{e}$ profiles shown in Figure \ref{fig:ThomsonScatteringProfiles}. While the original profiles are experimentally measured, the spline fits are constructed to approximate flux-surface-averaged quantities for use in the calculations. Ideally, these profiles would be true flux functions-- constant on equilibrium magnetic flux surfaces. However, when plotted in poloidal flux space, the measured profiles show significant deviations from flux surface constancy, with different values observed on the HFS and LFS of the same flux surface. This inconsistency may be related to the fact that, as shown in this work, a significant fraction of the edge region does not consist of closed, nested surfaces. In such regions, kinetic quantities would not be expected to align with flux surfaces, and the axisymmetric equilibrium itself becomes approximate. To provide consistent inputs for the modeling, a heavily smoothed spline fit was applied to each profile, enforcing a smoother and more flux-aligned structure; however, this remains an approximation. These are distinct from the lighter smoothing splines shown in Figure \ref{fig:ThomsonScatteringProfiles}, which do not represent flux-surface-averaged quantities.

\begin{figure}
\includegraphics[scale=0.8]{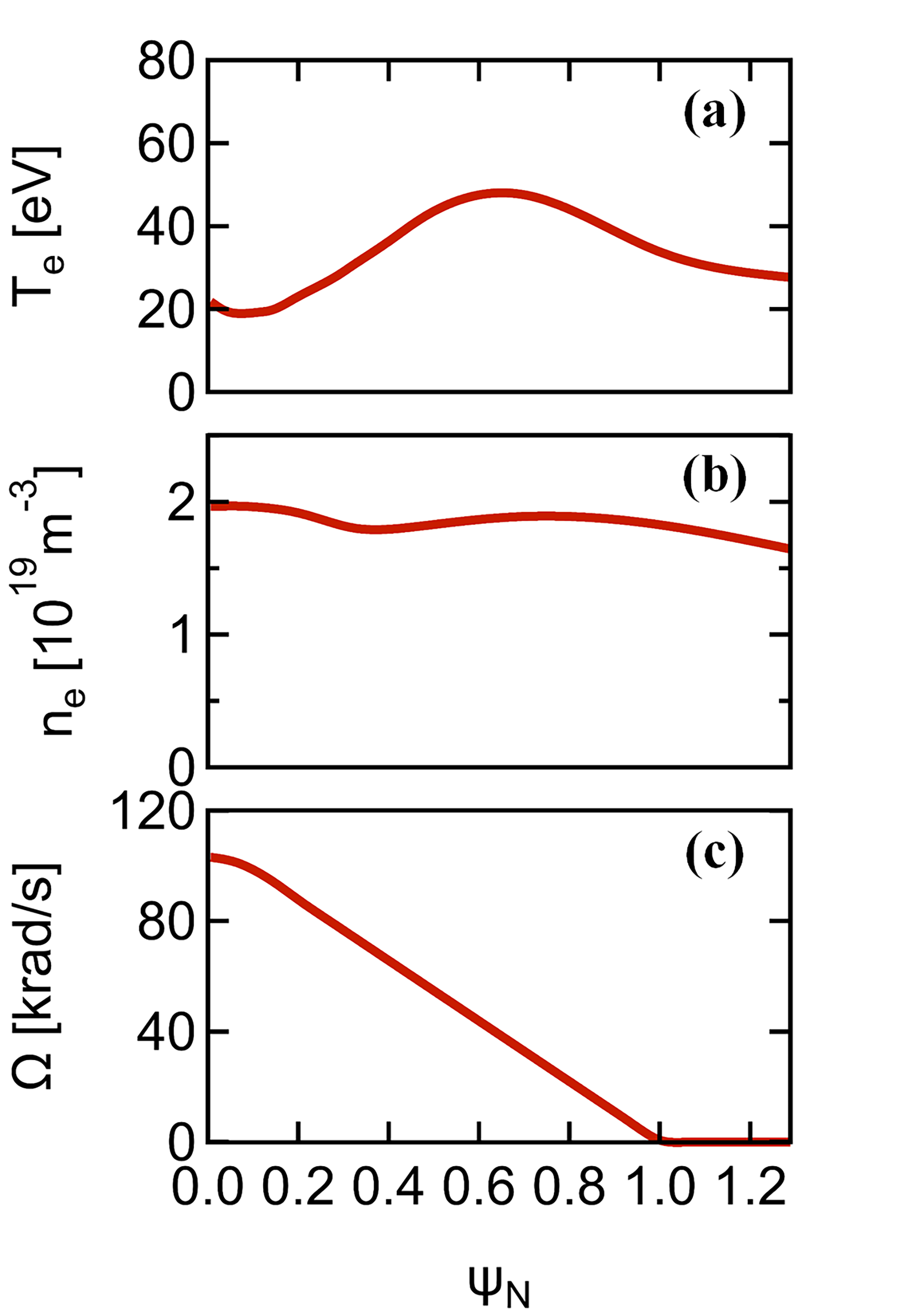}
\caption{\label{fig:KineticProfiles} Profiles used as inputs for the linear plasma response calculations: (a) smoothed, flux‐surface‐aligned spline fit of the measured $T_{e}$ profile; (b) corresponding $n_{e}$ profile produced using the same approach; and (c) assumed toroidal rotation profile, peaking at 100 krad/s in the core and decreasing linearly to zero at the LCFS.}
\end{figure}

The spline fit profiles also contain nonzero $T_{e}$ and $n_{e}$ values outside the last closed flux surface (LCFS). This feature is retained because the presence of the injected open field line current in this region makes nonzero values physically reasonable. The final $T_{e}$ and $n_{e}$ spline profiles used in the plasma response calculations are displayed in Figures \ref{fig:KineticProfiles}a and \ref{fig:KineticProfiles}b, respectively.

Additionally, since a plasma rotation profile is not currently available from measurements on Pegasus-III, a simple approximation of the toroidal angular rotation frequency was employed. The assumed profile peaks at 100 krad/s in the core and decreases linearly with $\Psi_{N}$, reaching zero at the LCFS. This profile is shown in Figure \ref{fig:KineticProfiles}c. The choice of 100 krad/s for the core rotation value is justified by several factors and represents a conservative upper bound, selected to maximize potential screening effects while remaining within the range suggested by existing studies and measurements. First, the core value is consistent with those used in previous M3D-C1 modeling studies of rotational screening effects \cite{Ferraro2012}, and was selected at the upper end of the scan range to reflect the possibility that current injection in LHI discharges contributes to enhanced plasma rotation via momentum injection. Measurements on Pegasus have shown toroidal rotation spanning from the tens to the low hundreds of krad/s; typical LHI plasmas exhibit rotation rates in the tens of krad/s, while high-energy ion outflows associated with filament reconnection events can reach speeds comparable to the local Alfvén velocity, corresponding to angular frequencies in the hundreds of krad/s range\cite{Burke2017}. For comparison, toroidal rotation velocities of $\sim$20 km/s have been measured in CHI discharges on NSTX, which correspond to angular frequencies on the order of $\sim$20 krad/s at the measurement location of $R=1.07$ m \cite{Nagata2007}. In the rotating cases presented here, the injected current stream is modeled as stationary in space--an approximation justified by its tie to the fixed injectors and reflected in the assumed rotation profile, which decreases to zero at the edge. A comparison case illustrating the resulting plasma response in the zero-rotation limit is presented later in this section.

\begin{figure}[t]
\includegraphics[width=\columnwidth]{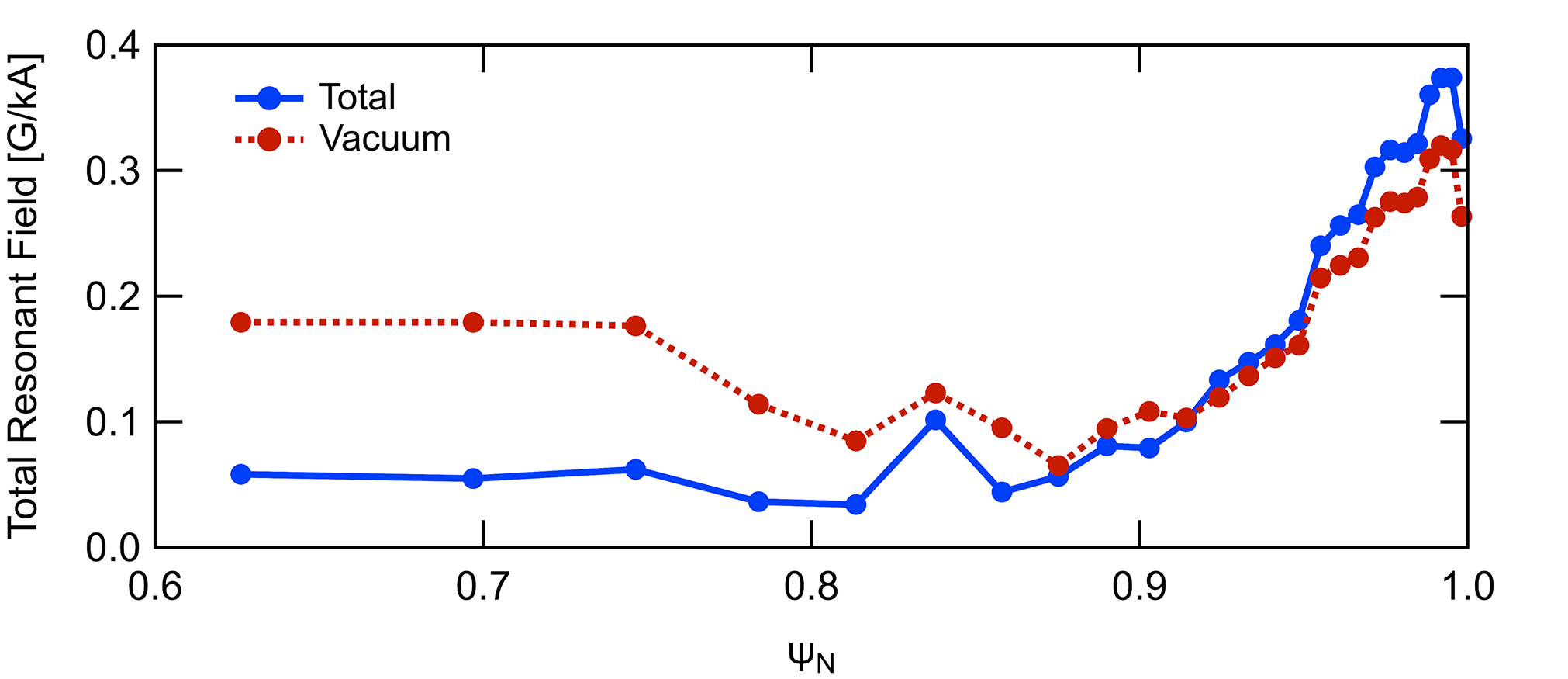}
\caption{\label{fig:Resonant_Field} Total resonant components of the “vacuum” magnetic field (red dashed) and the total field including the single‐fluid plasma response (blue solid) for the $n=1$ harmonic, shown versus normalized poloidal flux $\Psi_{N}$ for poloidal mode numbers $11 \leq m \leq 36$ (rational $q$ values in the same range).}
\end{figure}

Figure \ref{fig:Resonant_Field} shows the total resonant components of both the ``vacuum'' stream magnetic field and the total field including the single-fluid plasma response. These represent the amplitudes of the normal magnetic field perturbation at rational surfaces, evaluated in PEST coordinates for the $n=1$ harmonic. Resonant components are shown for poloidal mode numbers $m$ with $11 \leq m \leq 36$, corresponding to rational $q$ values in the same range. The absence of resonant components at normalized poloidal flux values $\Psi_{N} < 0.6$ reflects the structure of the $q$ profile. Sufficient spectral resolution was used to ensure accurate identification of resonances at integer $q$ values. Both the total and vacuum resonant fields exhibit a general increase with $\Psi_{N}$ beyond $\Psi_{N} \approx 0.9$, likely because the perturbation sources (the injected current streams) are physically closer to the outer plasma edge, enhancing coupling efficiency of the perturbation to nearby rational surfaces. Additionally, the vacuum field perturbation strength naturally decreases with distance from the source, which could contribute to weaker perturbations at surfaces deeper in the plasma. However, the spatial distribution of resonant field amplitudes is also influenced by the coupling of the perturbation to rational surfaces through their shared mode structure, so strong resonances might also be expected at surfaces deeper inside the plasma. This effect may explain the relatively higher vacuum resonant field values observed at the $q=11$–$13$ surfaces.

The plasma response reduces the amplitude of the resonant field at rational surfaces located farther radially inward (e.g., $q=11-15$), as evidenced by the total resonant field being a factor of $\sim$3 lower than the vacuum field in this region. This reduction is attributed to plasma screening effects. At surfaces farther radially outward, the total and vacuum field amplitudes are comparable. A modest amplification effect is observed near and beyond the $q=25$ surface, where the total field exceeds the vacuum field by $\sim$15$\%$. The reduction of the total resonant field relative to the vacuum resonant field at rational surfaces located farther inward may result from a combination of physical effects. One possibility is that increased plasma rotation toward the core enhances the generation of screening currents that oppose the applied perturbation, reducing the resonant field amplitude \cite{Fitzpatrick1995,Boozer1996,Ferraro2012}. These inner surfaces are also located in a hotter, lower-resistivity region of the plasma, which may further enhance the plasma’s ability to support screening currents.

\begin{figure}[b]
\includegraphics[width=\columnwidth]{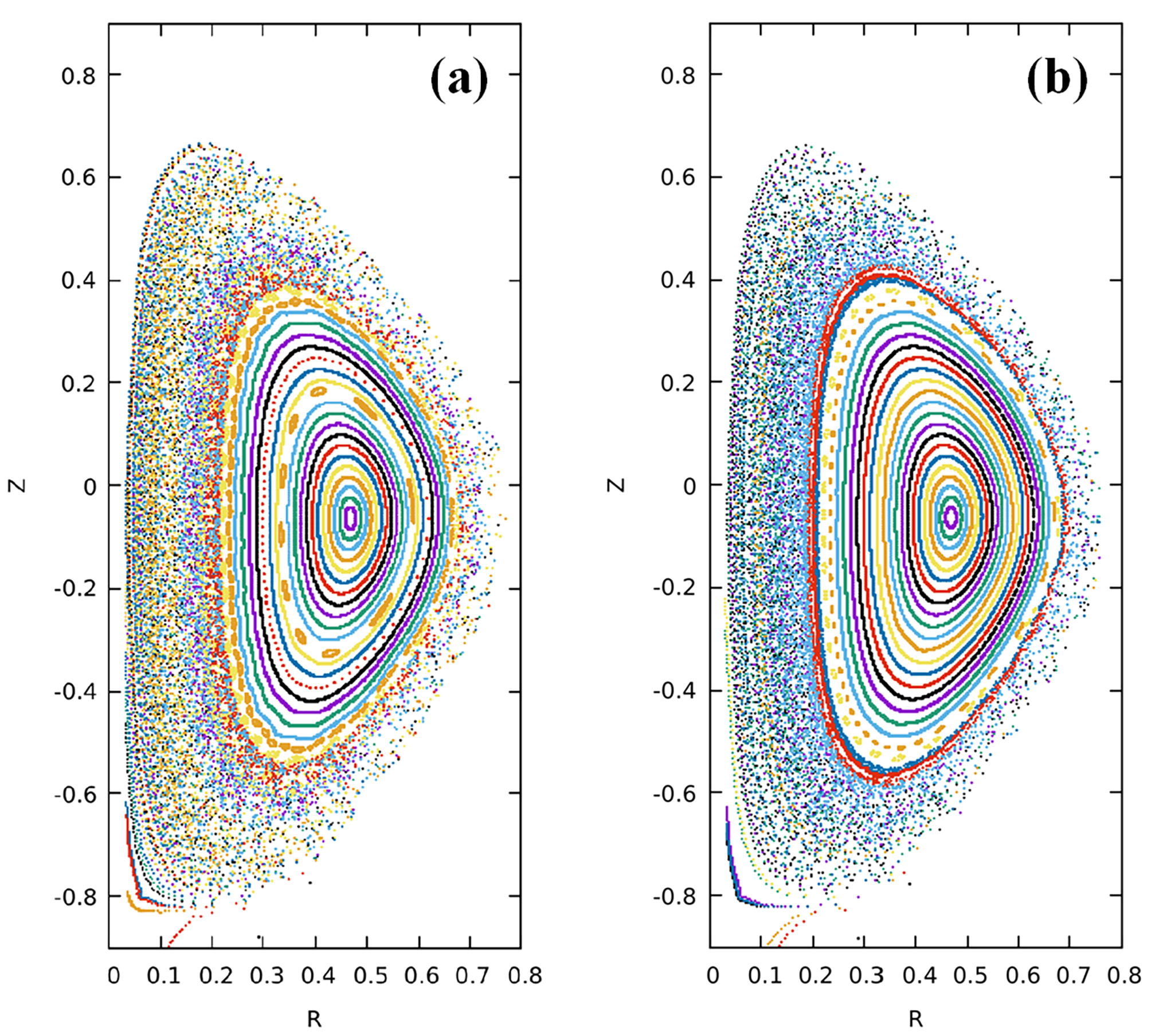}
\caption{\label{fig:Poincare_vac_and_total_compare} Poincaré plots in $(R,Z)$ space comparing magnetic field line structure for the vacuum field case (a) and the total field including the single‐fluid plasma response (b). In both cases, the Chirikov parameter exceeds unity at and beyond the $q=22$ surface ($\Psi_{N} \gtrsim 0.93$), indicating the onset of stochasticity, though deformation of flux surfaces is evident further inward.}
\end{figure}

\begin{figure}
\includegraphics[width=\columnwidth]{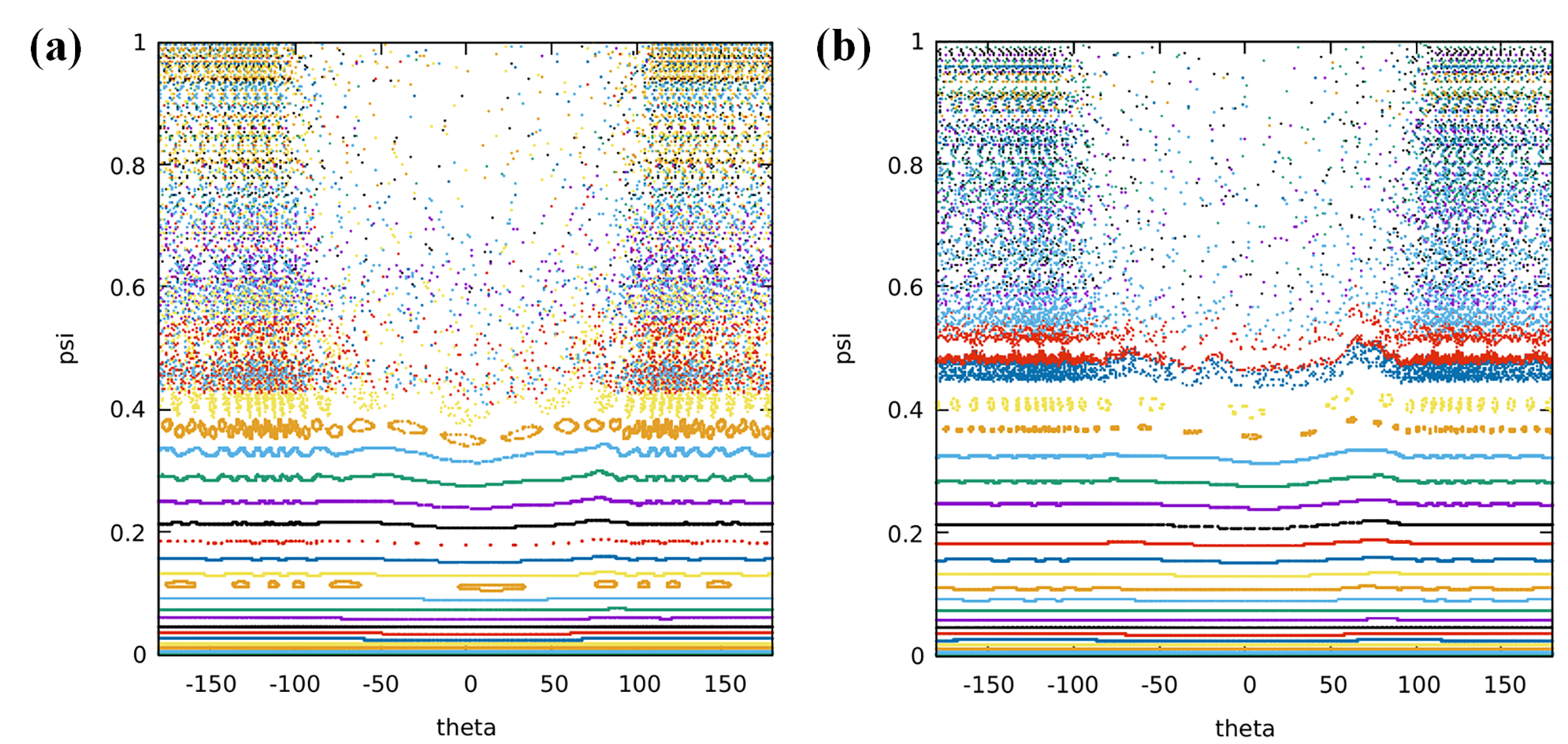}
\caption{\label{fig:Poincare_vac_and_total_compare_PsiTheta} Poincaré plots in $(\theta,\Psi_{N})$ space for the same vacuum field (a) and total field (b) cases as in Figure \ref{fig:Poincare_vac_and_total_compare}. Including the plasma response results in a modest reduction in island size and overlap relative to the vacuum field case.}
\end{figure}

Poincaré plots are used to visualize the magnetic field topology resulting from the stream perturbation, providing a qualitative assessment of the impact of the injected current streams on flux surface integrity. By tracing magnetic field lines and recording their intersections with a poloidal plane, these maps reveal whether magnetic surfaces remain closed, form islands, or become stochastic. Figures \ref{fig:Poincare_vac_and_total_compare} and \ref{fig:Poincare_vac_and_total_compare_PsiTheta} show in $R-Z$ and $\theta-\Psi_{N}$ space, respectively, the magnetic field structure of the perturbed plasma both without and including the plasma response. In these calculations, the total field is constructed by summing contributions from toroidal harmonics $n=1$ through $n=8$. It can be seen that in both cases, the stream induces a significant layer of chaotic fields. 

The Chirikov overlap criterion from nonlinear dynamics and chaos theory provides a quantitative threshold for the onset of chaos in systems with resonant perturbations \cite{Chirikov1979}. It predicts when neighboring magnetic island chains begin to overlap—specifically, when the sum of the half-widths of adjacent islands exceeds the distance between their centers. Once this overlap occurs, the regular nested magnetic surfaces are destroyed, and magnetic field lines follow chaotic trajectories. The Chirikov parameter measures this overlap; when it exceeds unity, stochasticity in the magnetic field is expected. In both cases shown in Figures \ref{fig:Poincare_vac_and_total_compare} and \ref{fig:Poincare_vac_and_total_compare_PsiTheta}, the point where the Chirikov parameter exceeds unity occurs at and beyond the $q=22$ surface, which corresponds to $\Psi_{N} \gtrsim 0.93$. However, the figures show a clear progression with increasing $\Psi_{N}$ at surfaces with $\Psi_{N} \lesssim 0.93$, as well. Flux surfaces closer to the core remain nested and closed. As $\Psi_{N}$ increases, surfaces become increasingly distorted, and magnetic islands eventually begin to emerge and grow in size. The onset of overlapping islands occurs at $0.4 \lesssim\Psi_{N} \lesssim0.6$, indicating that the perturbation significantly affects the magnetic topology even before the Chirikov parameter exceeds unity. Thus the transition from closed surfaces, to isolated island chains, and finally to overlapping islands and stochasticity is a gradual process that begins deeper within the plasma; significant magnetic field line perturbations, and the resulting enhancement of cross-filed transport that arises from parallel motion along chaotic field lines, can occur well before the formal onset of stochasticity as defined by the Chirikov criterion. 

The Chirikov parameter is also lower for surfaces with $\Psi_{N} \lesssim 0.92$ in the total field case that includes the plasma response, and this screening effect is clearly evident in the Poincaré plots themselves. It can be seen in Figures \ref{fig:Poincare_vac_and_total_compare} and \ref{fig:Poincare_vac_and_total_compare_PsiTheta} that a $q=10$ island chain at $\Psi_{N} \approx 0.12$ appears in the vacuum field case but is absent when the plasma response is included. Additionally, the vacuum field case exhibits more deformed surfaces at $0.2 \lesssim \Psi_{N} \lesssim 0.35$, a wider island chain at $\Psi_{N} \approx 0.37$, and partial island overlap occurring at lower $\Psi_{N}$ compared to the total field case. These differences are likely a result of the partial screening of the perturbation by the plasma response, which mitigates the formation and overlap of island chains. This further highlights the stabilizing role of the plasma response in preserving flux surface integrity in the presence of the LHI-induced perturbation. It also suggests that vacuum calculations may overestimate the extent and severity of island formation, stochasticity, and the associated transport in regions closer to the plasma core. 

\begin{figure}
\includegraphics[width=\columnwidth]{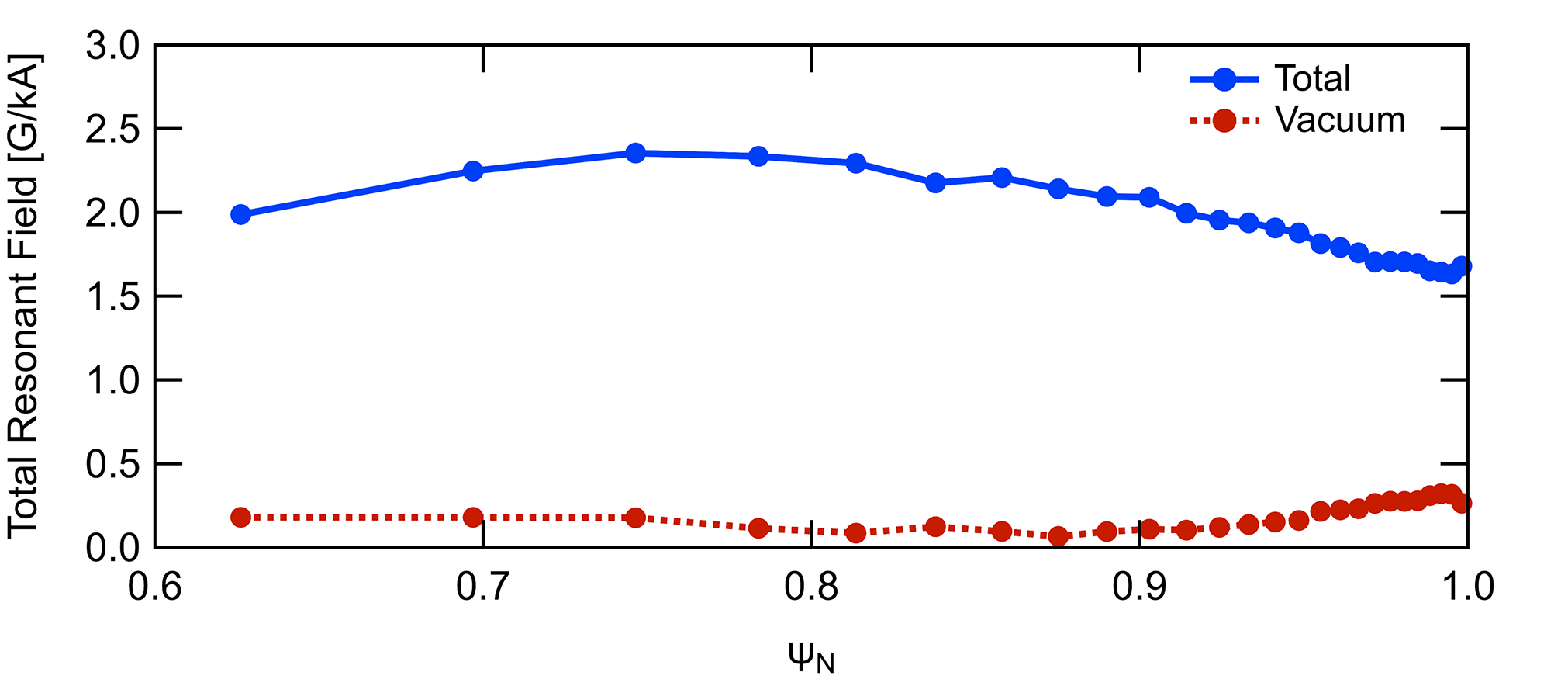}
\caption{\label{fig:Resonant_Field_ZeroRotation} Total resonant components of the “vacuum” magnetic field (red dashed) and the total field including the single‐fluid plasma response (blue solid) for the zero‐rotation case, shown versus normalized poloidal flux $\Psi_{N}$ for the $n=1$ harmonic and poloidal mode numbers $11 \leq m \leq 36$ (rational $q$ values in the same range). The total resonant field is about an order of magnitude larger than both the vacuum field and the corresponding rotating‐case total field, indicating strong amplification in the absence of rotational screening.}
\end{figure}

Plasma rotation is known to significantly affect the plasma response \cite{Fitzpatrick1995,Boozer1996,Ferraro2012}. To isolate this effect, we compare the case with prescribed rotation (using the kinetic profiles shown in Figure \ref{fig:KineticProfiles}) to the case in which the same $T_{e}$ and $n_{e}$ profiles are used, but the toroidal electron rotation is set to zero at all values of $\Psi_{N}$. Figure \ref{fig:Resonant_Field_ZeroRotation} shows the total resonant field, including the single-fluid plasma response, for the zero-rotation case, alongside the corresponding vacuum field, for the $n=1$ harmonic and poloidal modes $11 \leq m \leq 36$. The vacuum profile for this case is identical to the one shown in Figure \ref{fig:Resonant_Field}.

The total resonant field in the zero-rotation case is approximately an order of magnitude larger than its associated vacuum field, indicating strong amplification by the plasma. Compared to the rotating case, the total resonant field in the absence of rotation is also about an order of magnitude larger, underscoring the importance of rotation in mitigating the stream perturbation. These results are consistent with the expected role of rotational screening, whereby differential flow generates currents that counteract the imposed field. In the absence of rotation, the plasma response instead enhances the perturbation.


\begin{figure} [b]
\includegraphics[width=\columnwidth]{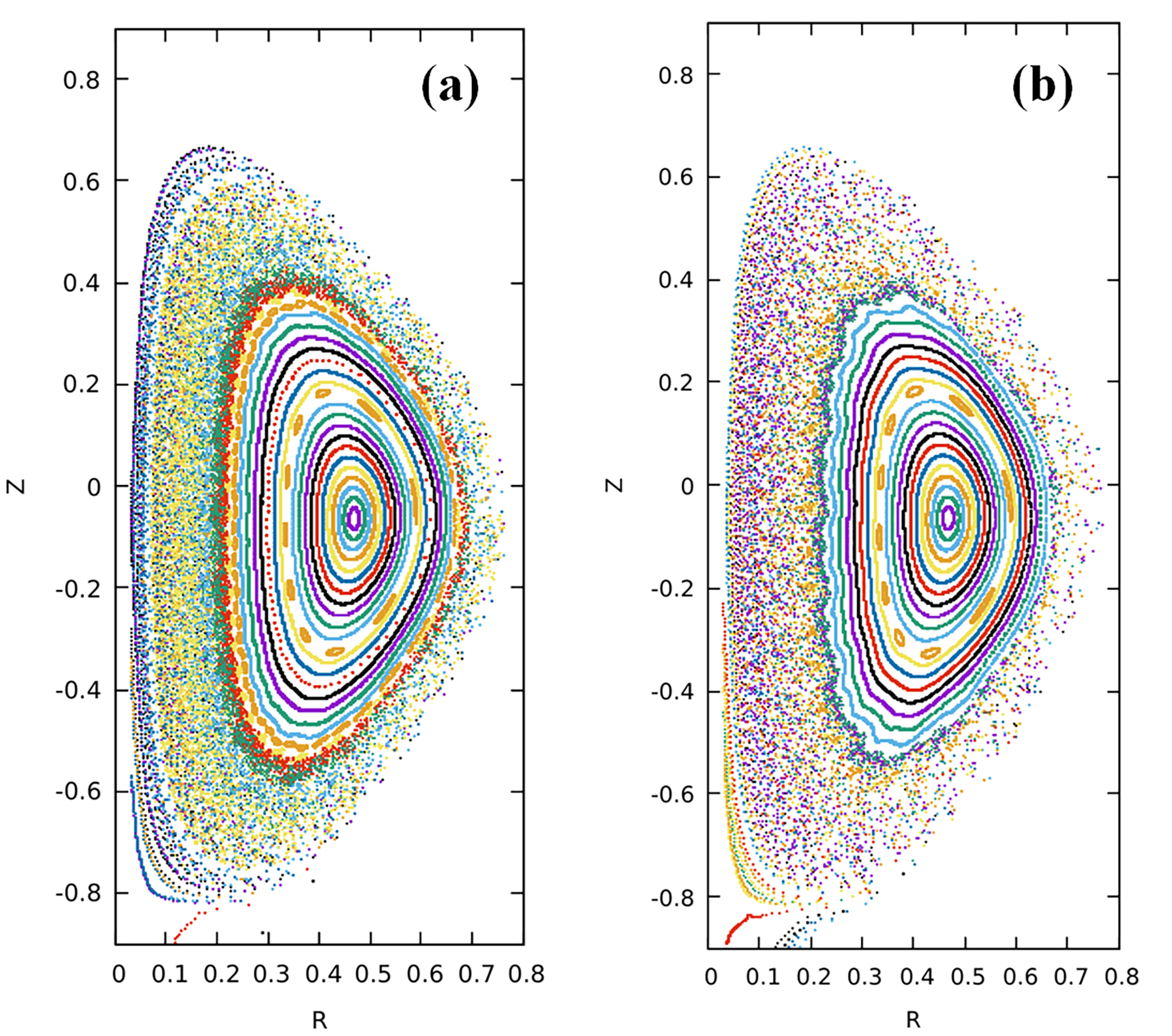}
\caption{\label{fig:Poincare_vac_and_total_compare_ZeroRot} Poincaré plots in $(R,Z)$ space comparing magnetic field line structure for the vacuum field case (a) and the total field including the single‐fluid plasma response (b) in the zero‐rotation case.}
\end{figure}

\begin{figure}
\includegraphics[width=\columnwidth]{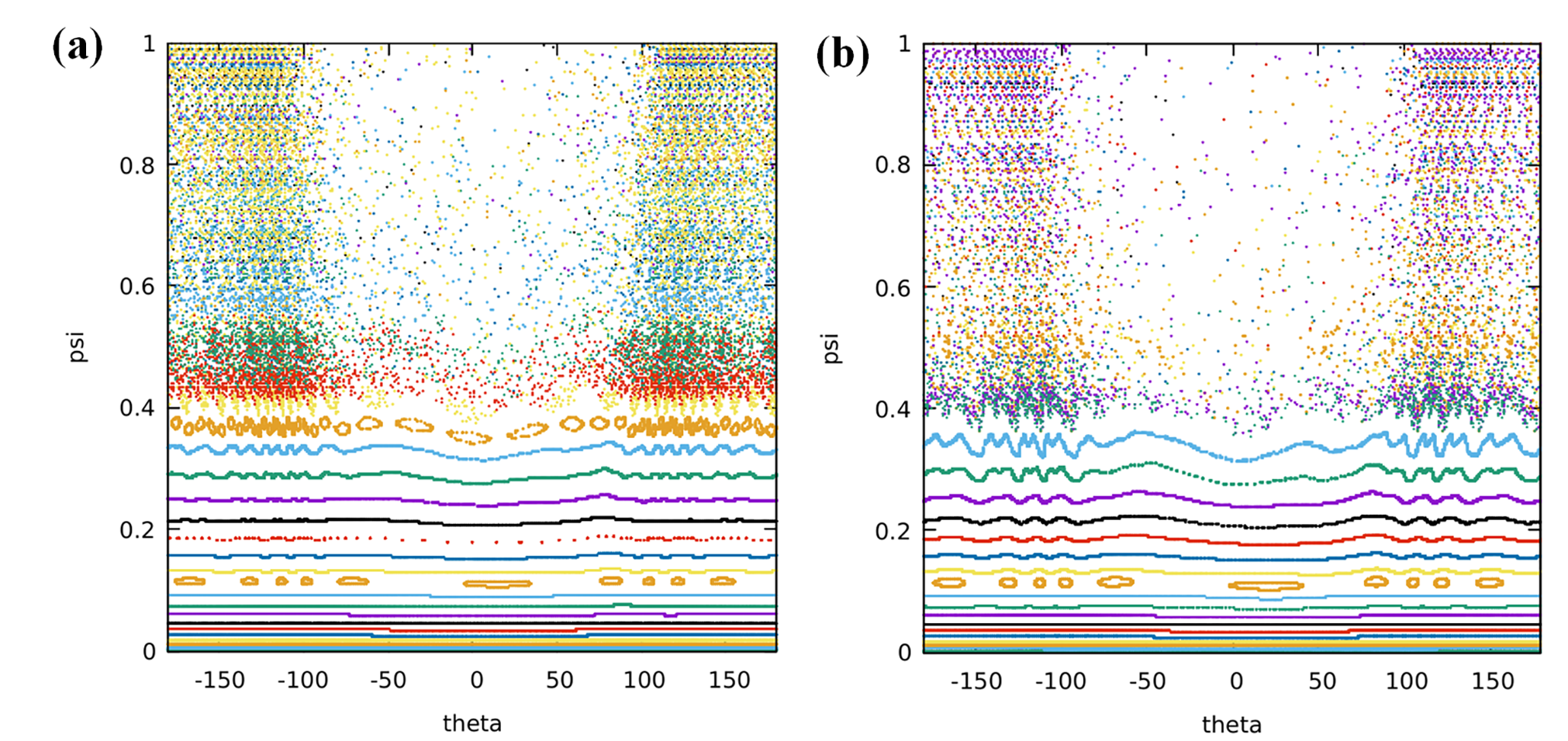}
\caption{\label{fig:Poincare_vac_and_total_compare_PsiTheta_ZeroRot} Poincaré plots in $(\theta,\Psi_{N})$ space for the same zero‐rotation vacuum field (a) and total field (b) cases as in Figure \ref{fig:Poincare_vac_and_total_compare_ZeroRot}. Without rotation, the plasma response enhances island formation and overlap relative to the vacuum field case, with deformation beginning deeper in the plasma than in the rotating case.}
\end{figure}

Figures \ref{fig:Poincare_vac_and_total_compare_ZeroRot} and \ref{fig:Poincare_vac_and_total_compare_PsiTheta_ZeroRot} show in $R-Z$ and $\theta-\Psi_{N}$ space, respectively, the magnetic field structure of the perturbed plasma, both without and including the plasma response, for the zero-rotation case. In these calculations, the total field is constructed by summing contributions from toroidal harmonics $n=1$ through $n=8$. Contrary to the rotating case, including the plasma response without rotation results in an onset of
surface deformation that occurs at lower values of $\Psi_{N}$. The Chirikov parameter, which again exceeds unity for the vacuum case at $\Psi_{N} \gtrsim 0.93$, is greater than unity across the entire $\Psi_{N}$ range plotted. Thus, the transition from closed surfaces to overlapping islands and stochasticity begins deeper within the plasma when the zero-rotation plasma response is included. This reinforces the role of plasma rotation in suppressing resonant amplification and maintaining flux surface integrity.

\subsection{Two-fluid}

Two-fluid effects are included here by setting the ion skin depth to its physical value. The resulting plasma response was first calculated using the same smooth profiles shown in Figure \ref{fig:KineticProfiles}. In this case, the rotation profile was assigned to be the electron rotation; M3D-C1 uses the prescribed profile to define an electron toroidal angular rotation frequency, from which the ion toroidal rotation frequency is determined using the pressure profile in the EQDSK file. Unlike the single-fluid case, where the ion, electron, and $\mathbf{E}\times\mathbf{B}$ rotations are all identical, introducing a finite ion skin depth allows for diamagnetic drifts that cause the ion and electron rotations to differ from one another and from the $\mathbf{E}\times\mathbf{B}$ rotation.

\begin{figure}
\includegraphics[width=\columnwidth]{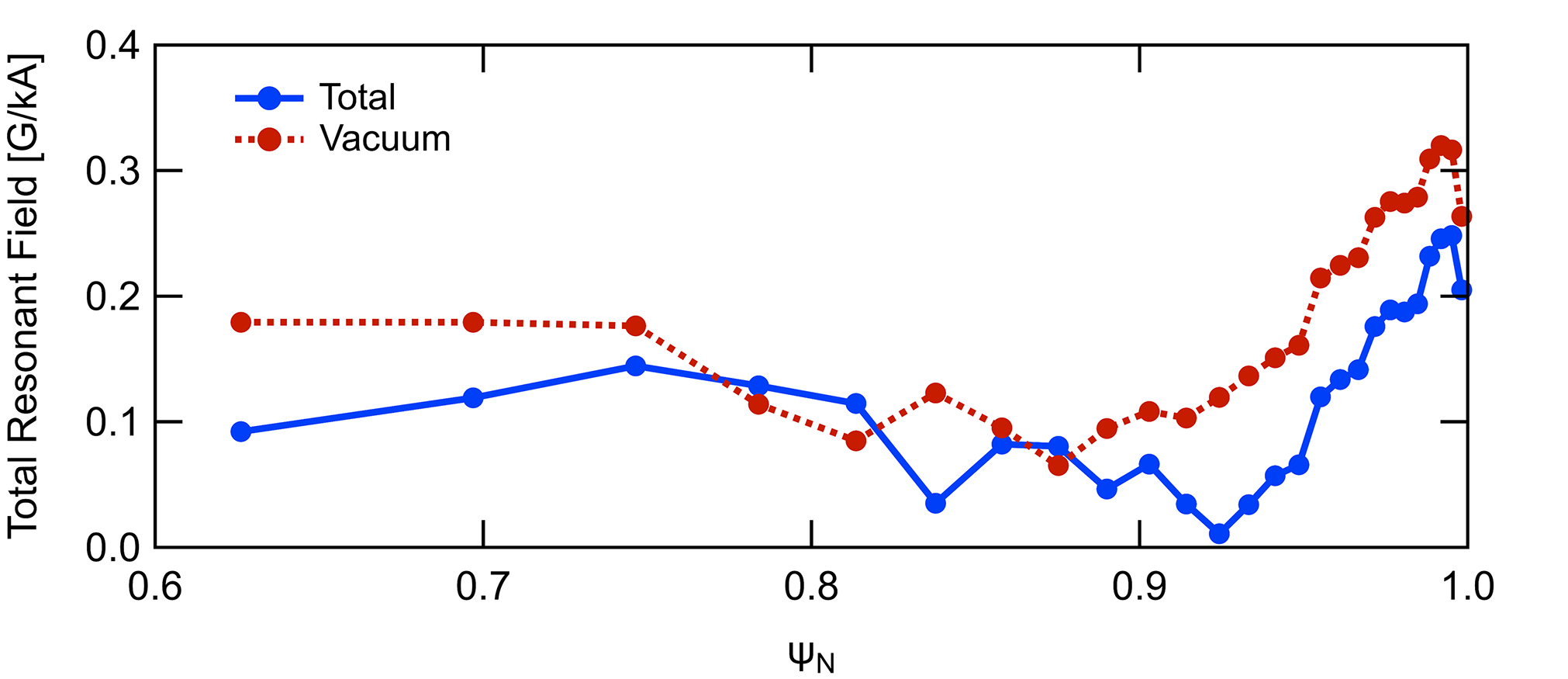}
\caption{\label{fig:Resonant_Field_TwoFluid} Total resonant components of the “vacuum” magnetic field (red dashed) and the total field including the two‐fluid plasma response (blue solid) for the $n=1$ harmonic, shown versus normalized poloidal flux $\Psi_{N}$ for poloidal mode numbers $11 \leq m \leq 36$ (rational $q$ values in the same range).}
\end{figure}

Figure \ref{fig:Resonant_Field_TwoFluid} shows the total resonant field, including the two-fluid plasma response, alongside the corresponding vacuum resonant field for the $n=1$ harmonic and poloidal modes $11 \leq m \leq 36$. This can be directly compared to the single-fluid case shown in Figure \ref{fig:Resonant_Field}. Notably, whereas the single-fluid response exhibited a modest amplification near and beyond the $q=25$ surface, the two-fluid response shows a clear screening effect for surfaces at and beyond the $q=19$ surface. The total resonant field is $\sim38\%$ lower than the vacuum field in this region. For surfaces with $q<19$, the plasma response continues to exhibit a shielding effect at most locations, though the reduction is generally weaker than in the single-fluid case. 

The enhanced screening observed near the edge in the two-fluid case may be attributed to the decoupling between ion and electron motion, which alters how the plasma supports currents. This decoupling enables the formation of finer-scale current structures and supports shorter-wavelength responses to counteract the applied perturbation \cite{Huba2003}. The magnetic geometry in this high $\mathrm{d}q/\mathrm{d}\Psi_{N}$ region near the edge is more responsive to such fine-scale perturbations, enhancing the effectiveness of the localized current response.

\begin{figure}[b]
\includegraphics[width=\columnwidth]{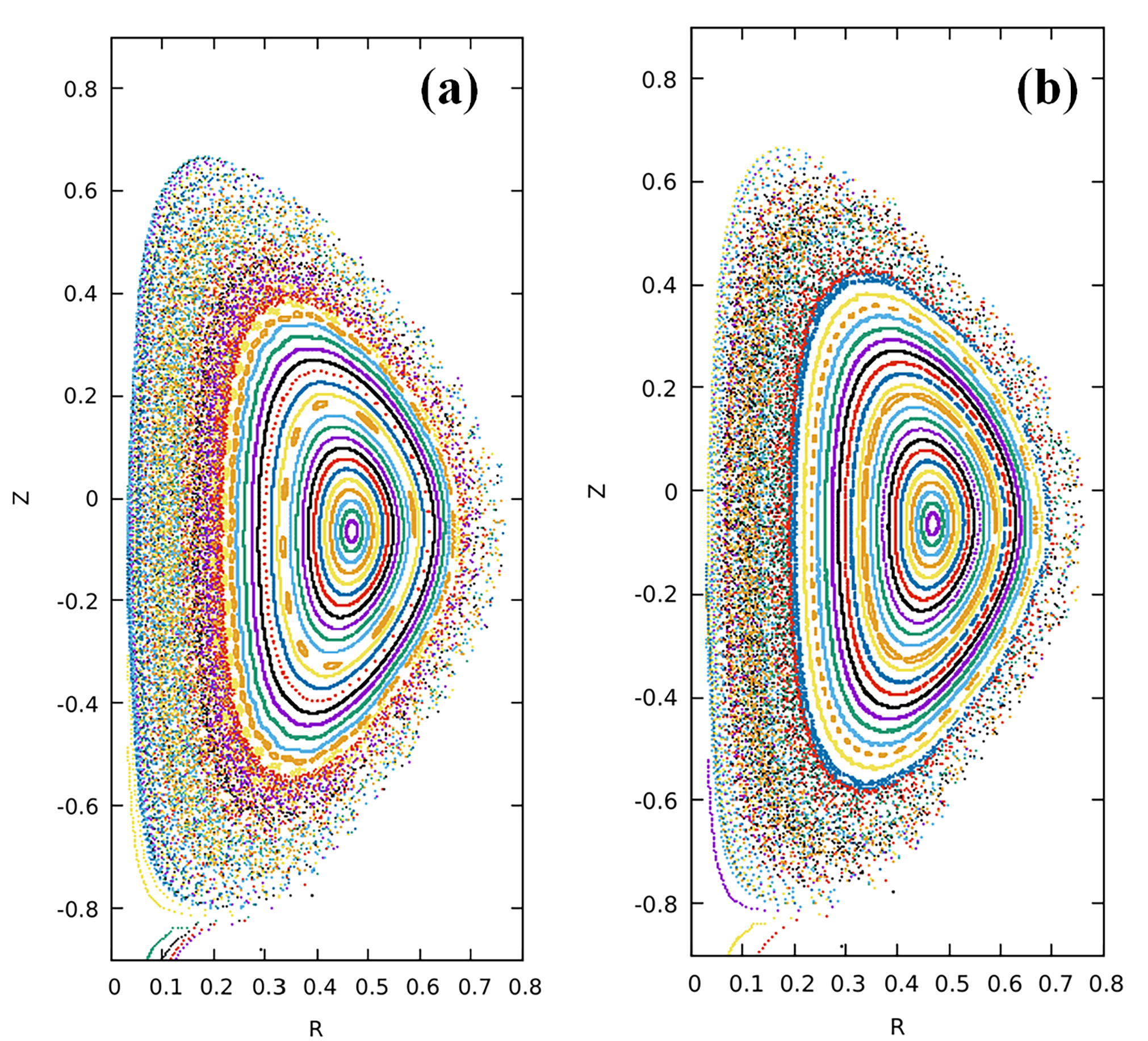}
\caption{\label{fig:Poincare_vac_and_total_compare_RZ_TwoFluid} Poincaré plots in $(R,Z)$ space comparing magnetic field line structure for the vacuum field case (a) and the total field including the two‐fluid plasma response (b).}
\end{figure}

\begin{figure}
\includegraphics[width=\columnwidth]{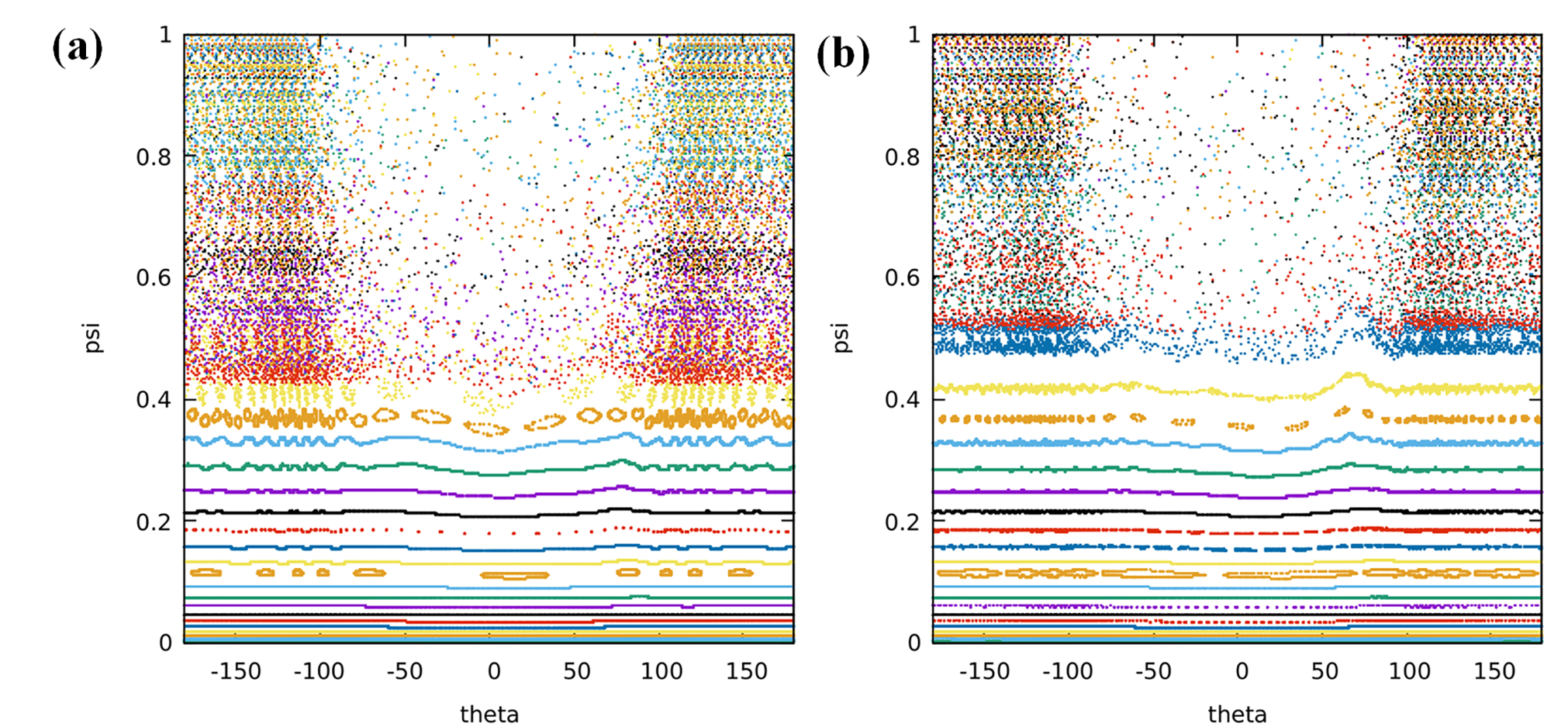}
\caption{\label{fig:Poincare_vac_and_total_compare_PsiTheta_TwoFluid} Poincaré plots in $(\theta,\Psi_{N})$ space for the same vacuum field (a) and total field (b) cases as in Figure \ref{fig:Poincare_vac_and_total_compare_RZ_TwoFluid}. Stronger edge screening is evident in the two‐fluid case, where the plasma response reduces inward field line chaos and island overlap, preserving surfaces such as the one at $\Psi_{N} \approx 0.41$ that are broken in the single‐fluid case.}
\end{figure}

The effects of this enhanced edge screening are evident in the Poincaré plots, as well. Figures \ref{fig:Poincare_vac_and_total_compare_RZ_TwoFluid} and \ref{fig:Poincare_vac_and_total_compare_PsiTheta_TwoFluid} show the magnetic field structure of the perturbed plasma in $R-Z$ and $\theta-\Psi_{N}$ space, respectively, both without and including the two-fluid plasma response. In these calculations, the total field is constructed by summing contributions from toroidal harmonics $n=1$ through $n=8$. Although screening is somewhat weaker at intermediate $q$ surfaces in the two-fluid case, the strong suppression of resonant fields near the edge appears to limit the inward propagation of stochasticity and island overlap, resulting in improved flux surface integrity deeper in the plasma. For example, comparing Figure \ref{fig:Poincare_vac_and_total_compare_PsiTheta_TwoFluid}b to its single-fluid equivalent in Figure \ref{fig:Poincare_vac_and_total_compare_PsiTheta}b reveals that the yellow island chain at $\Psi_{N} \approx 0.41$ present in the single-fluid case remains an intact surface when two-fluid effects are included. Furthermore, with the two-fluid plasma response, the Chirikov parameter does not exceed unity until the $q=25$ surface ($\Psi_{N} \gtrsim 0.95$), which is farther radially outward than in both the vacuum case and the single-fluid case.

\begin{figure}[h]
\includegraphics[width=\columnwidth]{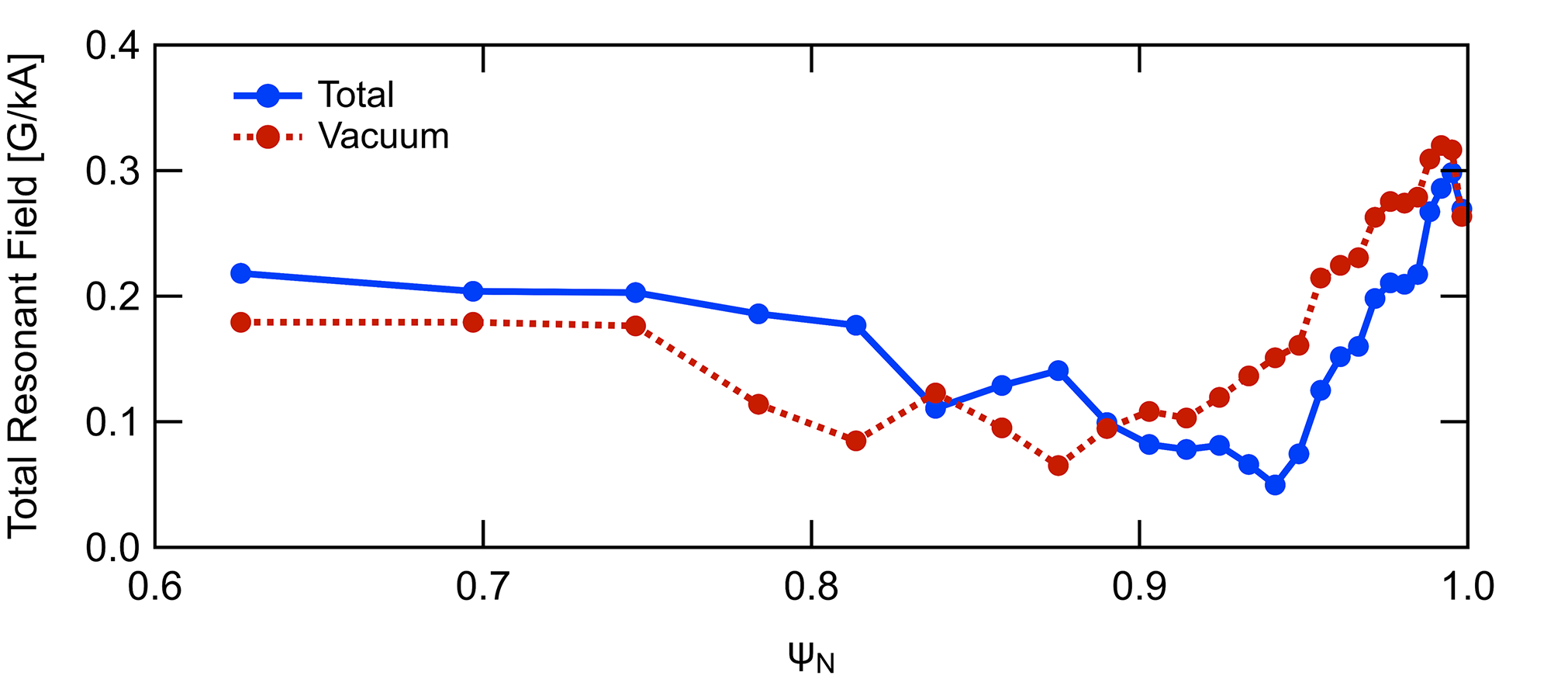}
\caption{\label{fig:Resonant_Field_ZeroRotation_TwoFluid} Total resonant components of the “vacuum” magnetic field (red dashed) and the total field including the two‐fluid plasma response (blue solid) for the zero‐rotation case, shown versus normalized poloidal flux $\Psi_{N}$ for the $n=1$ harmonic and poloidal mode numbers $11 \leq m \leq 36$ (rational $q$ values in the same range). Two-fluid effects significantly mitigate the amplification that is evident in the single-fluid, zero-rotation case, particularly near the edge where some screening is preserved even without rotation.}
\end{figure}

The enhanced screening enabled by two-fluid effects is further illustrated in the limit where the prescribed electron rotation is set to zero (noting that in the two-fluid case the ion and $\mathbf{E}\times\mathbf{B}$ rotations then differ self-consistently). Figure \ref{fig:Resonant_Field_ZeroRotation_TwoFluid} shows the total and vacuum resonant fields for the two-fluid case when the electron rotation profile is set to zero across all values of $\Psi_{N}$. As in the rotating case, the two-fluid plasma response exhibits a clear screening effect at and beyond the $q=20$ surface. In contrast, for surfaces with $q<20$, the total resonant field exceeds the vacuum field, indicating that rotation remains necessary to achieve screening in the plasma core. This result can be directly compared to the single-fluid, zero-rotation case shown in Figure \ref{fig:Resonant_Field_ZeroRotation}, where the absence of rotation leads to strong amplification of the resonant field across nearly the entire profile. In contrast, the two-fluid response significantly mitigates this amplification, particularly near the edge where some screening is preserved even without electron rotation. These results suggest that while rotation continues to play a critical role in screening, two-fluid physics substantially reduces the overall amplification of the resonant field in the limit where the prescribed electron rotation is set to zero. This highlights the importance of including two-fluid effects in modeling low-rotation scenarios, as single-fluid treatments may overpredict resonant field amplitudes and associated chaotic field line behavior. 

As discussed in Section \ref{subsec:Inputs}, hollow $T_{e}$ profiles have been observed in LHI discharges on Pegasus-III \cite{Tierney2024}. This naturally raises the question of whether the screening effect of the plasma response could reduce the extent of the chaotic edge layer if the $T_{e}$ profile were to fill in--that is, if the maximum temperature shown in Figure \ref{fig:KineticProfiles}a were sustained throughout the plasma core. This scenario might be reasonable if supplemental heating were introduced or if improved injector coupling efficiency led to the more peaked profiles previously observed on Pegasus \cite{Bodner2021}. In that sense, the assumed flat-core profile would represent a conservative case compared to a more strongly peaked temperature profile. 

The extent of the chaotic edge region remained largely unchanged when the $T_{e}$ profile was flattened across the core. Similarly, extending the profile to peak in the core produced negligible differences, suggesting that the plasma response is relatively insensitive to these moderate variations in core $T_{e}$ structure. It is possible that more substantial modifications to $T_{e}$ and other kinetic profiles--such as $n_{e}$ or $\Omega$--may have greater potential to influence the size and character of the chaotic region. It is also possible that the broad chaotic region arises in part from nonzero $T_{e}$ and $n_{e}$ values beyond the LCFS defined by the initial reconstructed equilibrium; however, these values are based on experimental measurements and were retained to preserve physical realism. This introduces a subtle complication: the LCFS is derived from an axisymmetric equilibrium that does not incorporate kinetic profile constraints. Incorporating measured kinetic profiles into future reconstructions may help resolve this ambiguity.

More broadly, understanding how kinetic profiles can be tailored to enhance flux surface integrity in the presence of injected current structures may offer a pathway to optimizing LHI performance. These results indicate that the toroidal rotation profile is a primary lever for maintaining flux surface integrity. Other factors such as overall MHD stability are also likely important but not considered here. The modeling shows that variations in rotation can qualitatively alter the degree of flux surface preservation. Accordingly, experimental measurements of toroidal flow velocity on Pegasus-III, from which the $\Omega$ profile can be inferred, would significantly enhance the fidelity of future modeling efforts.

\section{Refining Current Stream Model Using Edge $\bm{B}(R)$ Measurements}
\label{sec:Refine}

As discussed in Section \ref{sec:CurrentStreamModel}, the helical filament model employed to represent the injected current, while useful as a part of a first-order approach, is a simplification of a more complex current geometry. Previous work on Pegasus using an analysis of the phase and amplitude across a poloidal array of magnetic field pickup coils indicated that oscillatory motion of a discrete current stream in the edge region is the source of the $n=1$ magnetic activity described in Section \ref{sec:Relaxation} \cite{HinsonPhD2015,BarrPhD2016,Reusch2018}. Therefore, it is likely that the average stream geometry over an equilibrium timescale, which would include many cycles of the $n=1$ oscillation, would be more diffuse than a filament tied to a single magnetic field line. An analysis aimed at approximately quantifying this stream diffusivity using 3D edge $\bm{B}(R)$ measurements from an insertable Hall sensor probe\cite{Richner2018} is presented here.

\begin{figure}
\includegraphics[width=\columnwidth]
{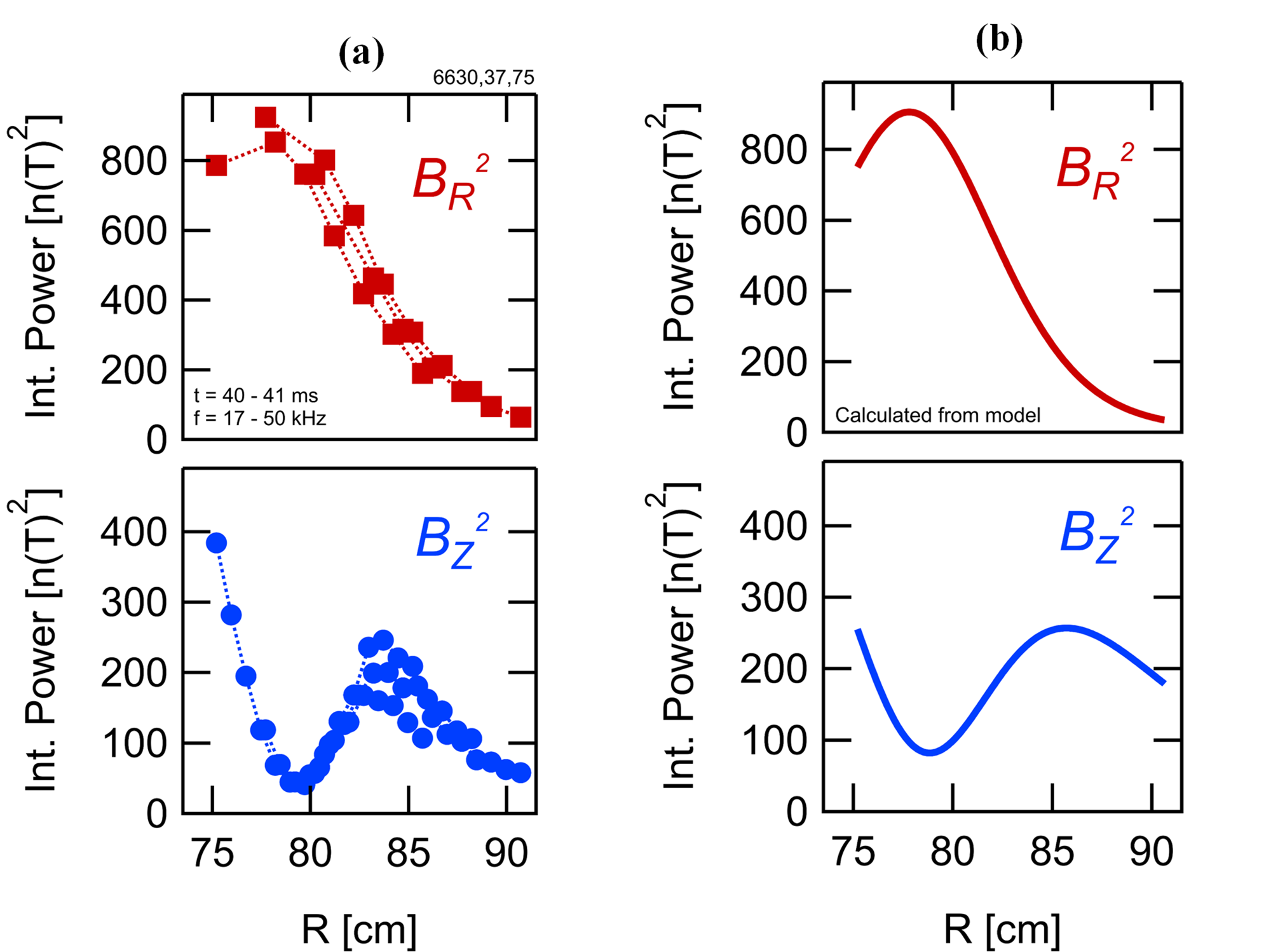}
\caption{\label{fig:Int_Power_Data_and_Calculated} (a) Measured radial profiles of magnetic field power in the $R$ (red) and $Z$ (blue) directions, integrated over the $n=1$ frequency band ($f=17 - 50$ kHz), from Hall probe measurements in three repeat discharges with a single injector array.
(b) Simulated magnetic power profiles from a model in which the injected current stream follows a toroidal loop trajectory with elliptical oscillation in the poloidal plane, showing good agreement with the measured $B_{R}$ and $B_{Z}$ profiles in both magnitude and shape.}
\end{figure}

The analyzed discharge was initiated with only one of the two injector arrays installed on Pegasus-III, ensuring that the majority of the magnetic field power measured by the probe could be attributed to the closest pass of the two streams (assumed to be co-located) from that single array. For these discharges, the $n=1$ mode was identified as the source of the magnetic power in the $f=17 - 50$ kHz range. Figure \ref{fig:Int_Power_Data_and_Calculated} shows radial profiles of the total magnetic field power in the $R$ and $Z$ directions integrated over the $n=1$ frequency range as measured with the Hall sensor probe from three repeat discharges. A 1 ms time window was used to generate the power spectra. The sensors have a sufficiently fast time response to resolve the $n=1$ mode, and isolating the frequency band associated with this feature reveals its radial position. The profiles show that the $B_{R}$ power peaks at $R \sim 77.5$ cm, while the $B_{Z}$ power has a local peak at $R \sim 83.5$ cm and a global maximum at an $R$ location radially inward of the innermost probe tip position. 

A simple model was developed that represents the current stream as a toroidal loop of current carrying the experimental total $I_{inj}$ parameterized to allow for transverse elliptical motion in the poloidal plane about a centroid. While this approximation ignores the pitch of the stream, it accounts for both the toroidal curvature of a helically winding stream as well as its oscillatory motion. The oscillation frequency was fixed to the $n=1$ frequency ($f=35$ kHz), which was taken to be the peak frequency observed in the power spectral density (PSD) of the probe Hall sensors. A Biot-Savart calculation was used to build the time-evolving Green's response in the $R$ and $Z$ directions from the oscillating toroidal loop of current to the $R$ range covered by the Hall sensor probe. The PSD of the response to this current source was simulated by taking the Fourier transform of the Green's response over a 1 ms time window, which would include many cycles of the $n=1$ oscillation. This PSD was then integrated over the same $n=1$ frequency range used to generate the profiles shown in Figure \ref{fig:Int_Power_Data_and_Calculated}a, thereby simulating the integrated magnetic field power in the $R$ and $Z$ directions measured by the probe.

The simulated profiles resulting from a stream with a centroid $\sim 15$ cm above the probe are shown in Figure \ref{fig:Int_Power_Data_and_Calculated}b. Given the pitch of the magnetic field in the LFS edge, it is reasonable to expect one toroidal pass of the stream located vertically above the probe in its poloidal plane. The elliptical path traced out by the stream in the poloidal plane that gives rise to the simulated profiles is shown in Figure \ref{fig:Int_Power_Data_and_Calculated}b. As shown, the vertical component of the oscillation is approximately three times larger than the radial one; the ellipse traced out by the stream has a semi-major axis of 6.0 cm and a semi-minor axis of 1.9 cm. This result has been observed previously \cite{BarrPhD2016} and was explained by the vertical motion resulting from the magnetic island coalescence instability observed in the NIMROD simulations, assuming this reconnection process continues after the initial relaxation. As described in Section \ref{sec:Relaxation}, this instability attracts adjacent passes of the stream prior to the large-scale reconnection events, which could result in a vertical stretching of the transverse oscillatory motion of the stream. However, at this time in the discharge, the field line pitch in the LFS edge is high enough such that only one pass of the stream is expected in the plane of the probe, assuming the injected current continues to follow the edge field line to some degree. Thus, the applicability of this explanation at this time in the discharge remains uncertain. Alternatively, the observed oscillatory motion may reflect a slower rotation of the filament with the bulk plasma, with periodic reconnection events accommodating the continuous winding of the stream due to its line-tying at the injector. In a time-averaged sense, such dynamics could likewise appear in the probe plane as stream spreading or oscillatory motion with an elliptical character.

\begin{figure} [t]
\includegraphics[scale=0.85]{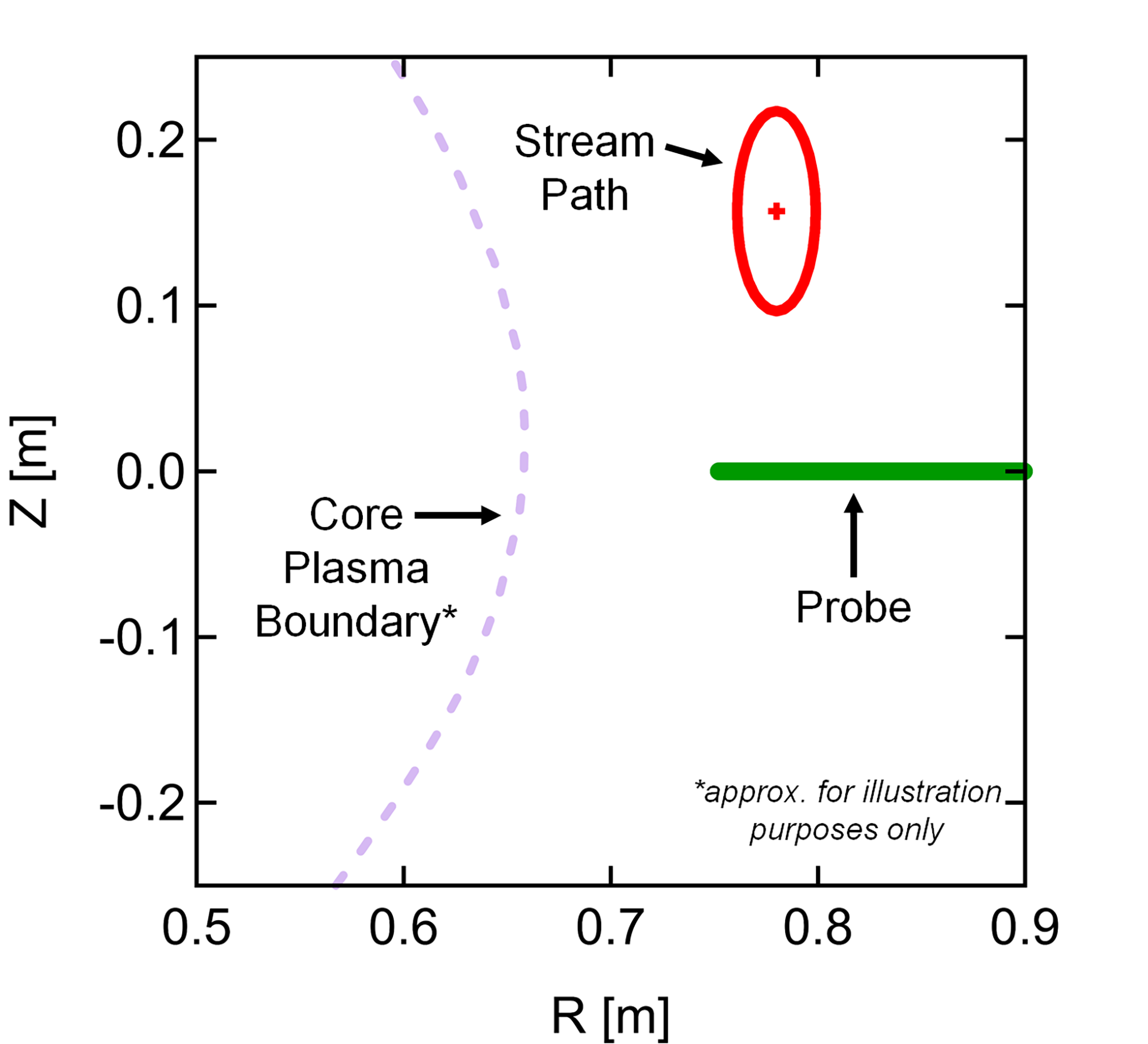}
\caption{\label{fig:Stream_Ellipse_Path_Annotated} Modeled oscillatory path of the injected current stream in the poloidal plane for the case in Figure \ref{fig:Int_Power_Data_and_Calculated}b, with a vertical semi‐axis of 6.0 cm and a radial semi‐axis of 1.9 cm. The centroid is located $\sim 15$ cm above the probe, consistent with an expected single toroidal pass in the probe plane.}
\end{figure}

Comparing the measured and simulated profiles in Figures \ref{fig:Int_Power_Data_and_Calculated}a and \ref{fig:Int_Power_Data_and_Calculated}b, respectively, indicates that the stream location and oscillation shown in Figure \ref{fig:Stream_Ellipse_Path_Annotated} reasonably reproduces the measurement-based profiles of magnetic power. The simulated $B(R)$ power profile matches the measured one in both magnitude and shape. The simulated $B(Z)$ power profile has a local peak that is wider and $\sim 2$ cm farther radially outward than the measured one, but has a very similar magnitude and shape otherwise. Notably, modeling the stream as a filament with little to no oscillation size (as described in Section \ref{sec:CurrentStreamModel}) was insufficient to reproduce the distinct profile shapes observed from the probe measurements. In fact, when the oscillation size is set to zero or even near zero, the predicted $B(R)$ and $B(Z)$ profiles are vanishingly small and essentially indistinguishable from zero on the scale of the measured data. These results, therefore, suggest that modeling the stream with spatial spreading, to represent distributed current and/or time-averaged oscillatory motion, offers a better match to experimental magnetic field measurements than a simple, rigid filament model. Although the underlying physical model remains approximate and the inferred geometry may not be unique, this approach provides a first-order means of incorporating edge magnetic measurements into a more realistic current stream representation. By adding a layer of physical complexity, it serves as an iterative improvement over the earlier helical filament approximation.

This result could be incorporated into future applications of the modeling described in Section \ref{sec:Modeling} by serving as an estimate of the stream spreading, which in turn could be used to generate a new applied 3D magnetic field for input to M3D-C1. For example, since the LHI stream structure is reasonably described as being fixed—or line-tied—to the injector face on one end and free to move on the other, it would be appropriate to construct a model that becomes increasingly diffuse with distance from the injector \cite{Lapenta2006,Reusch2018,OBryan2012,OBryan2014}. This spreading could be constrained using the size of the stream oscillation in the poloidal plane of the probe, as determined by this analysis. Comparing the results from the filament model to those that include some degree of stream spreading could provide insight into how sensitive the plasma response calculations--and subsequent magnetic structure modeling--are to spatial diffusivity and/or kinking motion of the current streams. 

\section{Conclusions}
\label{sec:Conclusions}

This work employed a simplified helical filament model of injected current streams in Pegasus-III LHI plasmas to compute their effect on magnetic topology. The M3D-C1 extended-MHD code was used to calculate the linear, time-independent plasma response to the 3D magnetic field generated by these modeled streams. Single-fluid and two-fluid plasma models, both with and without toroidal rotation, were systematically compared to assess their impact on resonant field amplitudes and magnetic surface deformation. Since the rotation profile proved to be a primary lever on flux-surface preservation, improved experimental constraints on toroidal flow would substantially enhance the fidelity of future modeling efforts.

In all cases examined, including those that showed some degree of screening by the response, Poincaré mapping of field lines--including contributions from harmonics $n=1$ through $n=8$--revealed substantial degradation of magnetic flux surfaces. The innermost appearance of an island chain that transitions outward into overlapping structures and widespread surface distortion occurred near $\Psi_{N} \approx 0.37$,  marking the beginning of a region extending toward the edge where flux surfaces become increasingly deformed.  While the full extent of this region does not exhibit fully stochastic behavior by the Chirikov criterion, it nonetheless represents a zone characterized by significant magnetic perturbation.

In rotating plasmas, both single-fluid and two-fluid models exhibited partial screening of the $n=1$ perturbation across portions of the profile, with similar reductions in island size and overlap at intermediate $q$ surfaces. However, the two-fluid model produced stronger suppression of the $n=1$ resonant fields near the plasma edge. 

In the absence of rotation, the single-fluid response strongly amplified resonant fields across the profile, driving island overlap and stochasticity well inside the edge. In contrast, the two-fluid case with zero prescribed electron rotation preserved much of the edge screening seen in the rotating case, while surfaces closer to the core experienced some amplification relative to the vacuum calculation--though still less than in the single-fluid, non-rotating case. These results confirm that rotation is critical for broad screening, while also showing that two-fluid physics can mitigate amplification under low-rotation conditions, particularly near the edge.

An analysis using magnetic probe measurements further indicated that modeling the stream with spatial spreading, potentially representing oscillatory motion, reproduces measured magnetic power profiles more closely than a rigid filament model. This suggests that refining the applied field geometry could improve predictive accuracy.

Future work should directly compare modeled magnetic fields and Poincaré plots to spatially and temporally resolved probe measurements. Such comparisons, absent in this study, would help validate both the filament and diffuse-stream models, constrain geometric parameters, and assess the sensitivity of plasma response predictions to stream structure. Incorporating these validation steps will be essential for developing higher-fidelity 3D field models that capture the complex, dynamic nature of injected current streams.

\begin{acknowledgments}
The authors thank M.W. Bongard for contributing the KFIT equilibrium reconstruction, T.N. Tierney and J.A. Reusch for contributing the Thomson scattering measurements, N.J. Richner for probe diagnostic development, and the Pegasus-III team for operating and maintaining the facility. This material is based on work supported by the U.S. Department of Energy, Office of Science, Office of Fusion Energy Sciences, under award no. DE-SC0019008. Any opinions, findings, and conclusions or recommendations expressed in this paper are those of the authors and do not necessarily reflect the views of the U.S. Department of Energy. 
\end{acknowledgments}

\section*{Author Declarations}
\subsection*{Conflict of Interest}
The authors have no conflicts to disclose.

\subsection*{Author Contributions}
\textbf{C. E. Schaefer:} Conceptualization (equal); Data curation (equal); Formal analysis (lead); Methodology (equal); Software (supporting); Visualization (lead); Writing -- original draft preparation (lead); Writing -- review \& editing (equal). \textbf{A. C. Sontag:} Conceptualization (equal); Formal analysis (supporting); Methodology (equal); Software (supporting); Writing -- original draft preparation (supporting); Writing -- review \& editing (equal). \textbf{N. M. Ferraro:} Conceptualization (equal); Data curation (equal); Formal analysis (supporting); Methodology (equal); Resources (lead); Software (lead); Writing -- review \& editing (equal). \textbf{J. D. Weberski:} Conceptualization (equal); Formal analysis (supporting); Methodology (equal); Software (supporting); Writing -- review \& editing (equal).  \textbf{S. J. Diem:} Conceptualization (equal); Formal analysis (supporting); Funding acquisition (lead); Methodology (equal); Project administration (lead); Supervision (lead); Writing -- original draft preparation (supporting); Writing -- review \& editing (equal).

\section*{Data Availability Statement}

Data from this publication are publicly available in openly documented, machine-readable formats \cite{DataAvail}. Note: reference and DOI accompanying the manuscript's dataset will be activated and a final ID assigned if the paper is accepted for publication.

\appendix

\section{Resonant Field Amplitudes for n=1-8 Across All Modeled Cases}
\label{sec:appendix}

For completeness, this appendix presents the resonant field amplitudes for toroidal mode numbers $n=1$–8 across all four modeled cases: single-fluid with and without rotation, and two-fluid with and without rotation. These spectra illustrate the range of resonant responses obtained in the linear calculations, including variations in amplitude and radial structure with harmonic number. While detailed comparisons in the main text focus on $n=1$, owing to its experimental accessibility and physical relevance, the higher-$n$ spectra provide important context: they contribute to the total field structure used in the Poincaré plots and highlight that no single harmonic fully characterizes the plasma response.

\begin{figure} [h]
\includegraphics[width=\columnwidth]{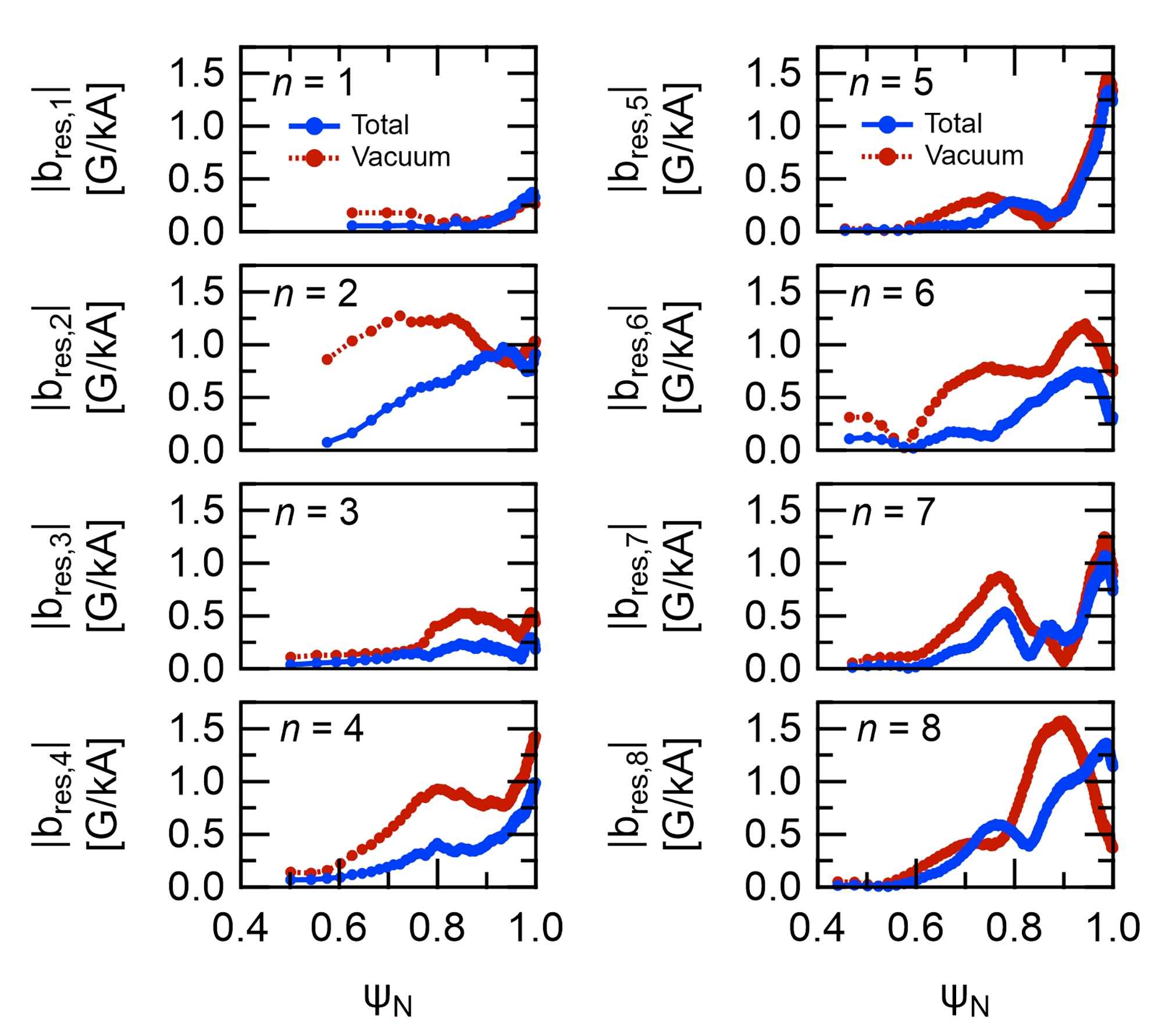}
\caption{\label{fig:SingleFluid_withRotation_AllN} Total resonant components of the “vacuum” magnetic field (red dashed) and the total field including the single-fluid plasma response with rotation (blue solid) for toroidal harmonics $n=1-8$, shown versus normalized poloidal flux $\Psi_{N}$. The differing numbers of points in each profile reflect the distinct sets of rational $q=m/n$ surfaces that are accessed as $n$ varies.}
\end{figure}

\begin{figure} 
\includegraphics[width=\columnwidth]{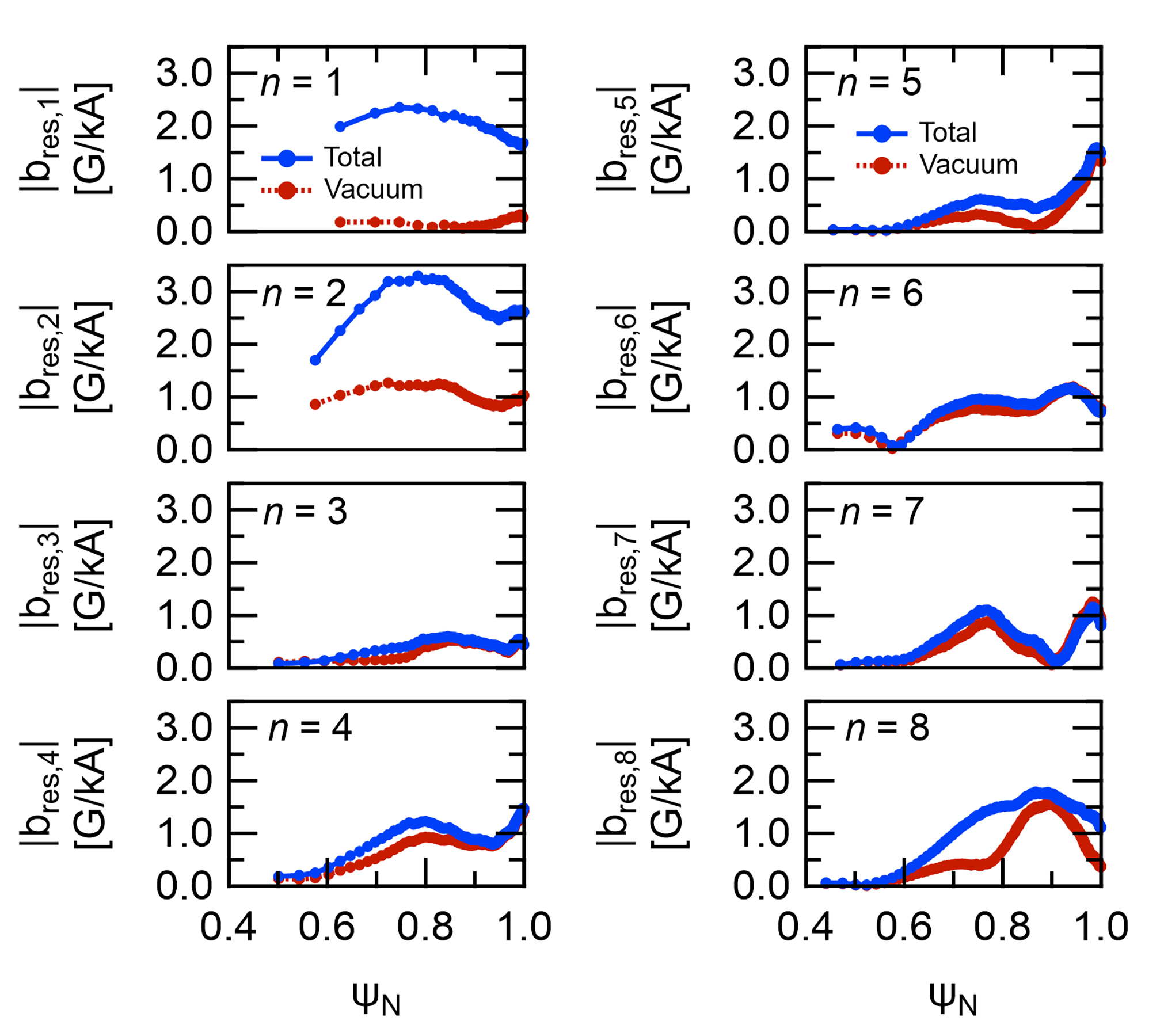}
\caption{\label{fig:SingleFluid_NoRotation_AllN} Total resonant components of the “vacuum” magnetic field (red dashed) and the total field including the single-fluid plasma response in the zero electron rotation case (blue solid) for toroidal harmonics $n=1-8$, shown versus $\Psi_{N}$. Note that the vertical scale differs from that in Figure \ref{fig:SingleFluid_withRotation_AllN}, reflecting the stronger amplification that occurs without rotational screening at some surfaces.}
\end{figure}

\begin{figure}
\includegraphics[width=\columnwidth]{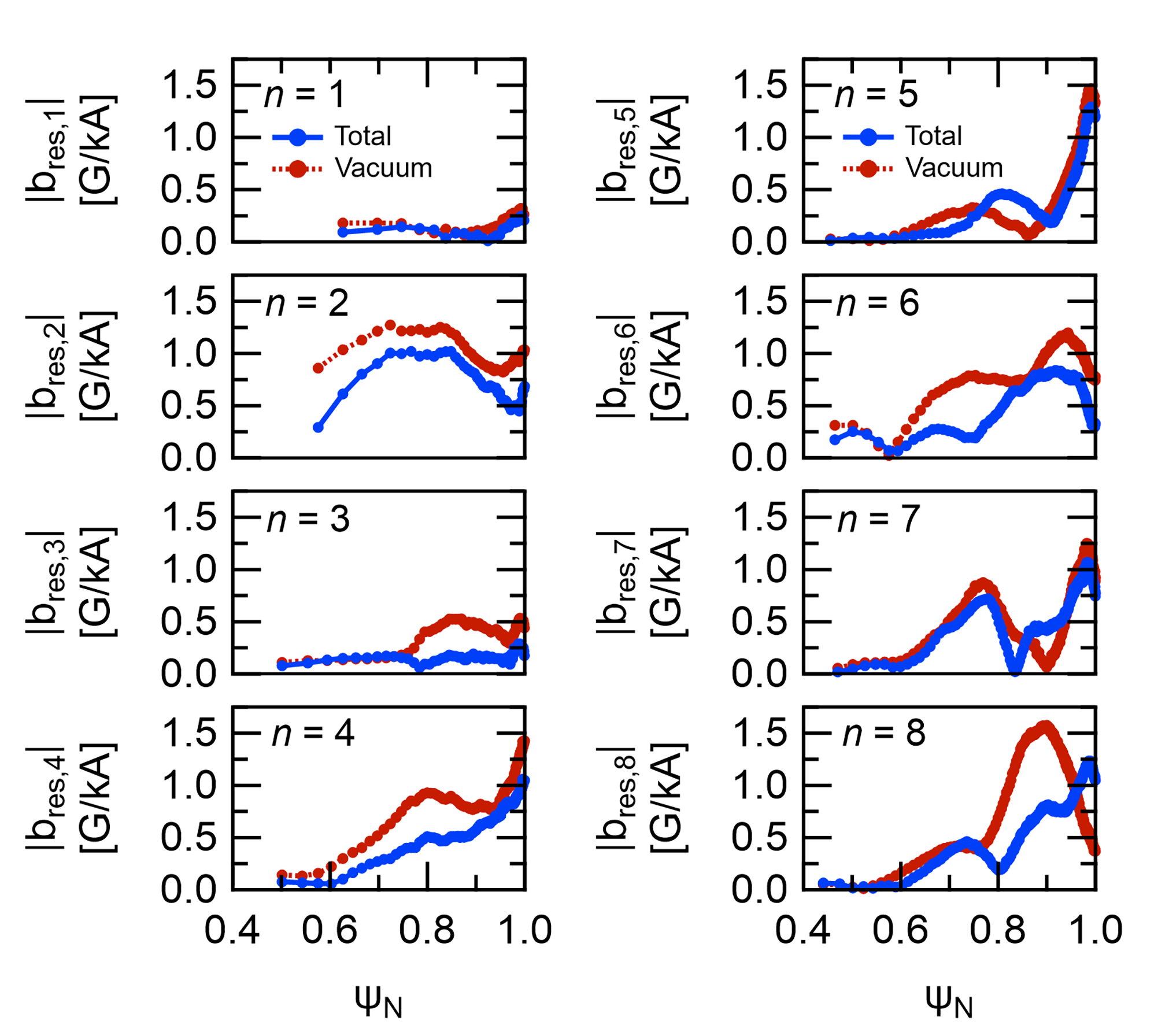}
\caption{\label{fig:TwoFluid_withRotation_AllN}  Total resonant components of the “vacuum” magnetic field (red dashed) and the total field including the two-fluid plasma response with rotation (blue solid) for toroidal harmonics $n=1-8$, shown versus $\Psi_{N}$. Note that the vertical scale is the same as in Figure \ref{fig:SingleFluid_withRotation_AllN}.}
\end{figure}

\begin{figure}
\includegraphics[width=\columnwidth]{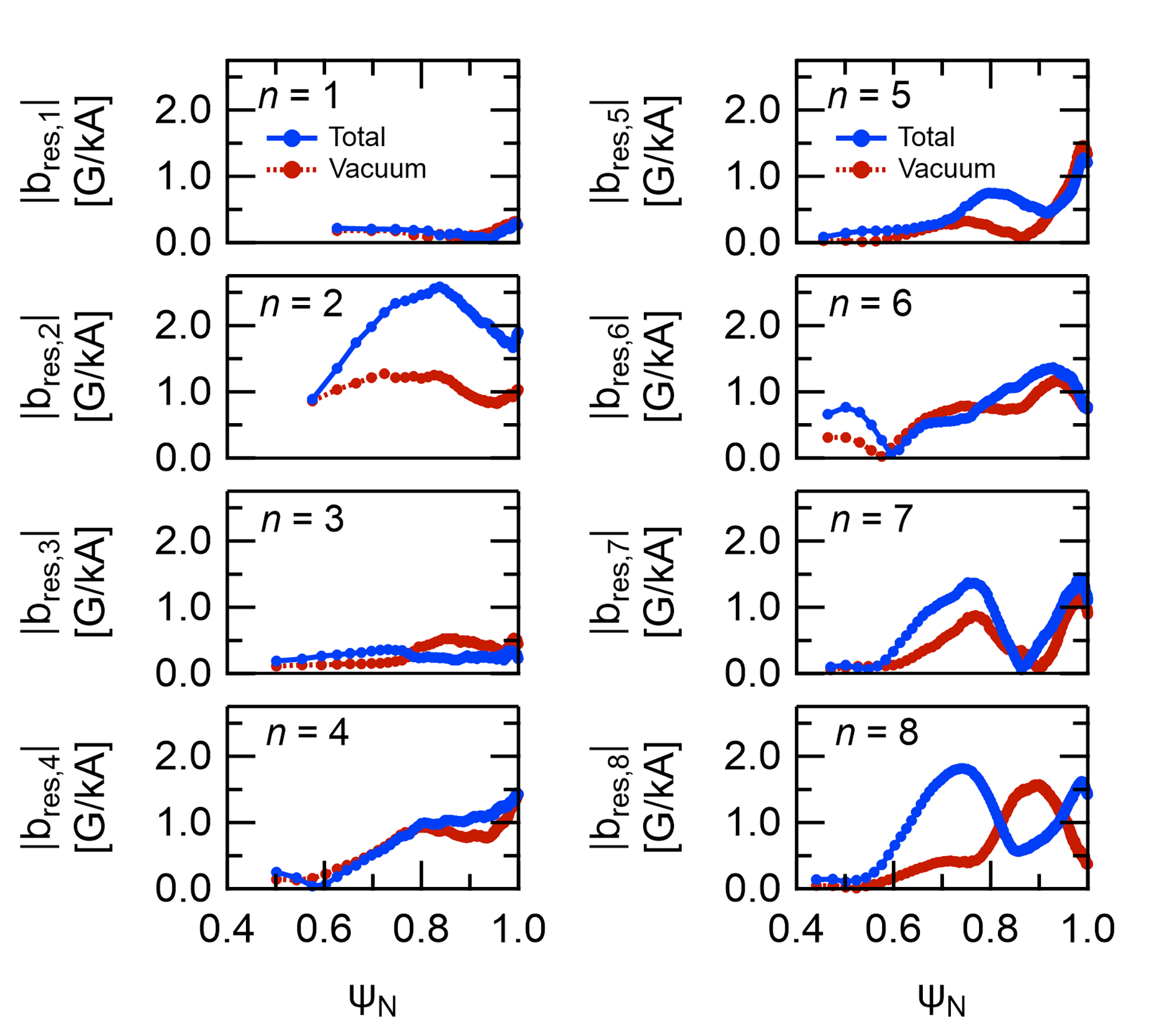}
\caption{\label{fig:TwoFluid_NoRotation_AllN} Total resonant components of the “vacuum” magnetic field (red dashed) and the total field including the two-fluid plasma response in the zero electron rotation case (blue solid) for toroidal harmonics $n=1-8$, shown versus $\Psi_{N}$. Note that the vertical scale differs from that in Figures \ref{fig:SingleFluid_withRotation_AllN}, \ref{fig:SingleFluid_NoRotation_AllN}, and \ref{fig:TwoFluid_withRotation_AllN}.}
\end{figure}

\FloatBarrier


\section*{References}
\bibliographystyle{aipnum4-2}
\bibliography{aipsamp}

\end{document}